\documentclass[final,5p,times,twocolumn]{elsarticle}

\usepackage{xr}
\makeatletter

\newcommand*{\addFileDependency}[1]{
\typeout{(#1)}
\@addtofilelist{#1}
\IfFileExists{#1}{}{\typeout{No file #1.}}
}\makeatother

\usepackage{amsmath,amssymb,graphicx,booktabs,subcaption,url}
\usepackage{enumitem}
\usepackage[table]{xcolor}
\usepackage{hyperref}
\hypersetup{
  hidelinks,
  breaklinks=true,
  pdfkeywords={Neural image compression, Discrete cosine transform, Spectral analysis, Frequency response}
}
\usepackage[final]{microtype}
\usepackage{amsthm}

\usepackage{caption}
\AtBeginDocument{%
  \setlength{\abovedisplayskip}{6pt plus 2pt minus 2pt}%
  \setlength{\belowdisplayskip}{6pt plus 2pt minus 2pt}%
  \setlength{\abovedisplayshortskip}{4pt plus 2pt minus 1pt}%
  \setlength{\belowdisplayshortskip}{6pt plus 2pt minus 2pt}%
}
\usepackage{adjustbox}
\usepackage{listings}

\theoremstyle{definition}

\theoremstyle{remark}

\usepackage{multirow} 
\usepackage{placeins}

\graphicspath{{images/}}
\journal{Neurocomputing}
\biboptions{sort&compress}

\begin{document}
\begin{frontmatter}

\title{Benchmarking Neural Image Compression with DCT-Based Spectral Diagnostics}

\author[skoltech]{Nikolay I. Kalmykov}
\ead{nikolay.kalmykov@skoltech.ru}
\author[skoltech]{Razan Dibo}
\author[uestc]{Yujun Du}
\author[uestc]{Yipeng Liu}
\author[skoltech,axxx]{Ivan Oseledets}
\author[skoltech]{Anh-Huy Phan\corref{cor1}}
\ead{a.phan@skoltech.ru}
\cortext[cor1]{Corresponding author}

\address[skoltech]{Skolkovo Institute of Science and Technology (Skoltech), Moscow, Russia}
\address[uestc]{University of Electronic Science and Technology of China (UESTC), Chengdu, China}
\address[axxx]{AIRI, Moscow, Russia}

\begin{abstract}
Neural image compression (NIC) models are commonly evaluated with scalar rate-distortion metrics, which do not show how a codec treats individual spatial frequencies. We propose a benchmark that probes codecs with the DCT basis matrix and single-frequency test images. We measure relative off-target energy (leakage), the fraction of reconstruction energy outside the probed frequency, together with target-frequency gain. Exact decompositions and bounds relate these measurements to reconstruction error. At operating points nearest 1\,bpp, the BMSHJ, MBT, and Cheng2020 models leave 80-97\% of their reconstructed high-band energy outside the target bin and have target-frequency gains of $0.02$-$0.23$, whereas TCM and FTIC attain off-target energy in the classical-codec range. Both probe types identify behavior absent from scalar metrics: released Cheng2020-Anchor models produce extreme pre-clamp outputs on isolated single-frequency inputs. Finally, decoder-only fine-tuning with natural-image and leakage losses reduces basis off-target energy by 90-99.8\% across all eight evaluated architectures while keeping the bitstream fixed, at a Kodak PSNR cost of $0.09$-$0.59$\,dB.

\par\medskip\noindent\textbf{Code} - \url{https://github.com/nkalmykovsk/dct_benchmark_nic}
\end{abstract}


\end{frontmatter}

\section{Introduction}
\label{sec:intro}

\paragraph{Motivation and problem setting}
Neural image compression (NIC) models now match or surpass traditional codecs on standard benchmarks such as Kodak~\cite{kodakdataset}, CLIC~\cite{clicdataset}, and DIV2K~\cite{div2kdataset}. Their performance is typically summarized by scalar metrics such as PSNR and MS-SSIM, which measure overall fidelity but provide little insight into how individual image structures are preserved~\cite{Yang2022An, ma2019image}. As a result, two codecs with similar rate-distortion performance may allocate compression resources very differently across spatial frequencies~\cite{lieberman2023spectral}.

Most NIC frameworks~\cite{balle2017endtoend, cheng2020attention, Liu_2023_CVPR} provide multiple ``quality'' variants corresponding to different rate-distortion trade-offs~\cite{mentzer2020hific, he2022elic, Chen2020Variable}. These variants are identified by scalar quality indices or nominal bitrate targets, which do not reveal how they differ in preserving textures, edges, and fine detail~\cite{wang2004ssim, blau2019rdp}. Although high-frequency attenuation in learned codecs has been observed qualitatively~\cite{mentzer2020hific}, systematic measurements of its magnitude, frequency dependence, and variation across codec architectures remain limited.

This motivates two questions:

\begin{enumerate}
\item How do codec architectures and quality variants allocate information across different frequency bands?

\item Do higher-quality settings genuinely preserve more high-frequency detail, or merely improve global rate-distortion scores?
\end{enumerate}

\begin{figure}[t]
\centering
\begin{subfigure}{0.32\linewidth}
    \centering
    \includegraphics[width=\linewidth]{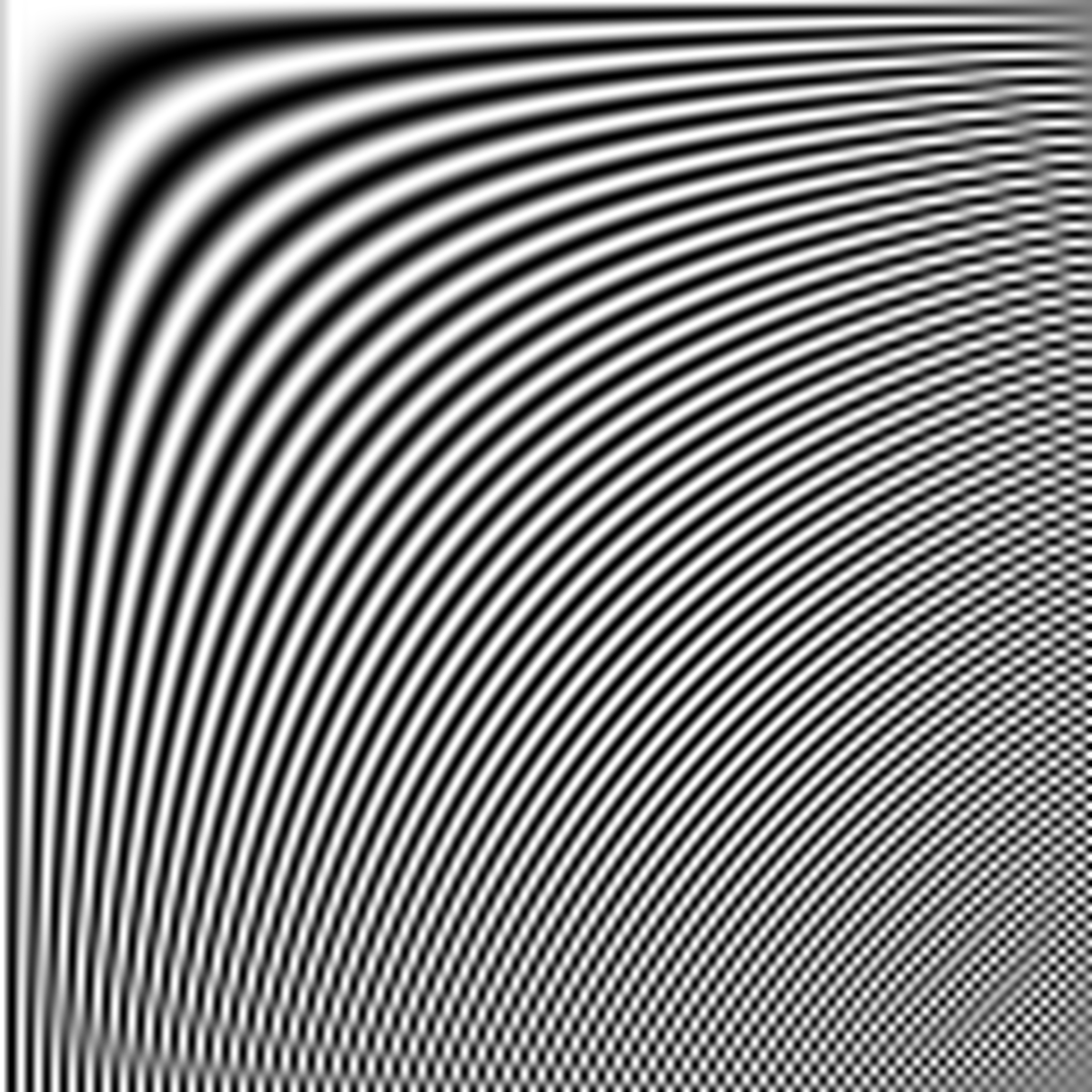}
    \caption{DCT basis matrix.}
    \label{fig:dct_orig_rgb}
\end{subfigure} \hfill
\begin{subfigure}{0.32\linewidth}
    \centering
    \includegraphics[width=\linewidth]{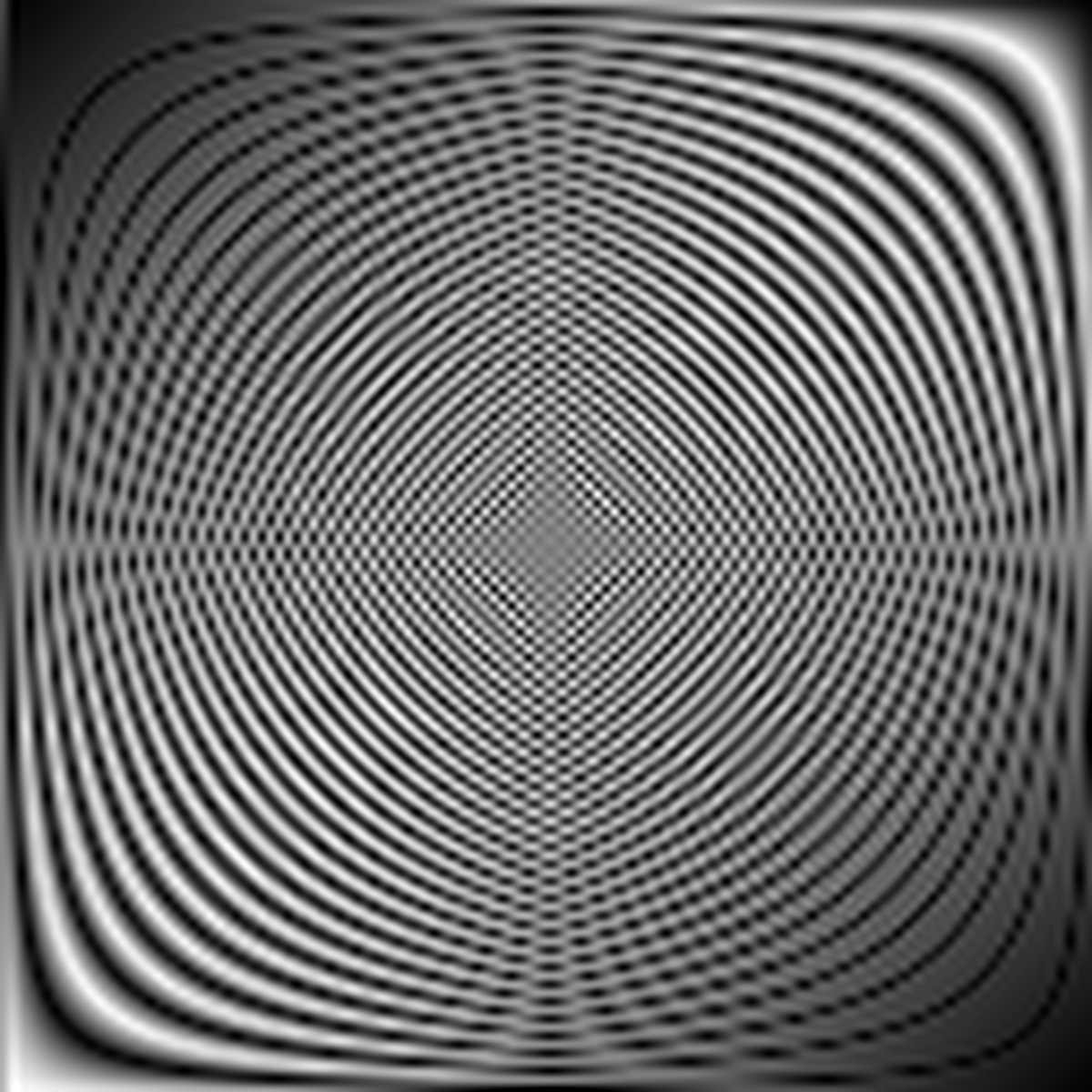}
    \caption{Frequency map.}
    \label{fig:dct_anchor}
\end{subfigure} \hfill
\begin{subfigure}{0.32\linewidth}
    \centering
    \includegraphics[width=\linewidth]{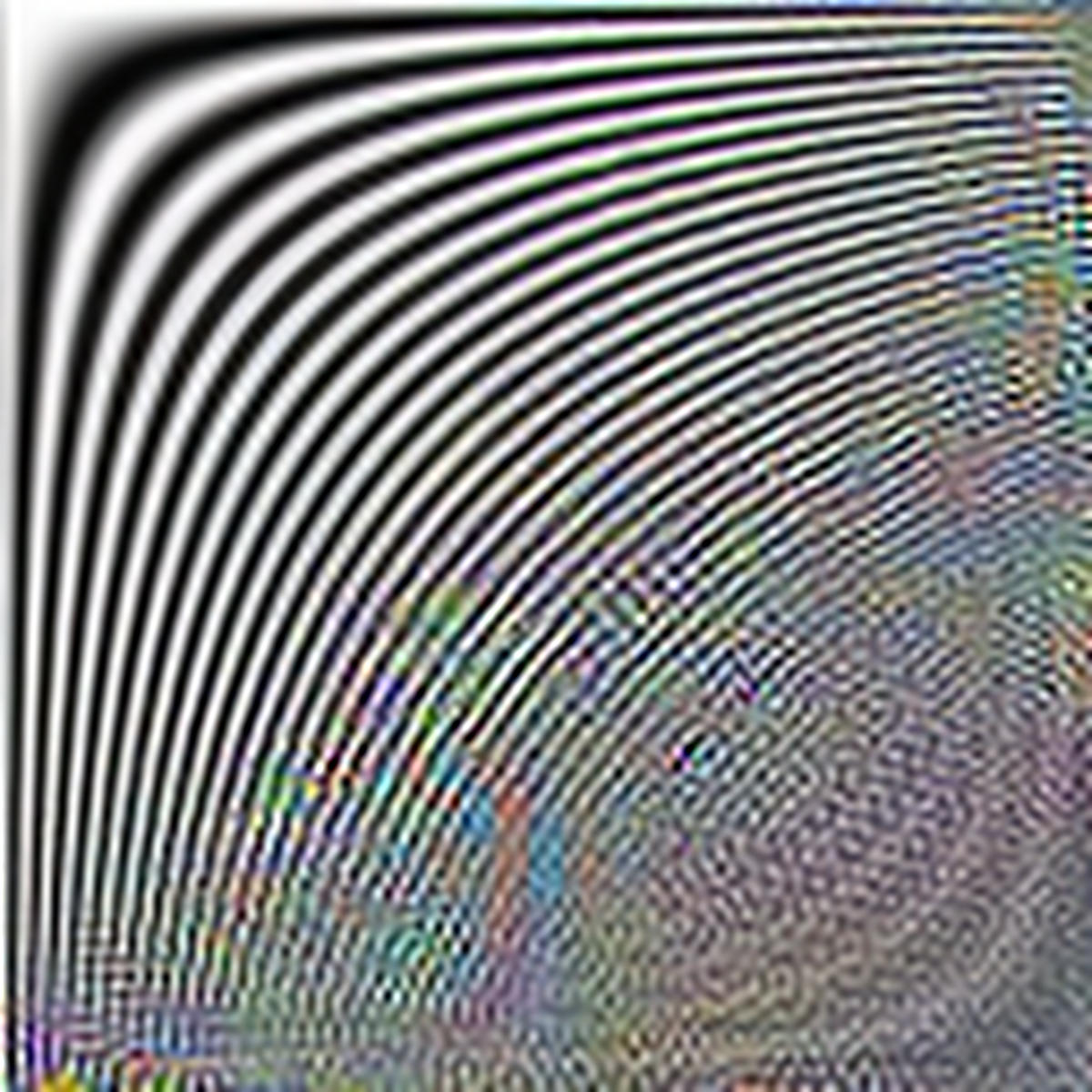}
    \caption{Cheng2020-Anchor.}
    \label{fig:dct_map}
\end{subfigure}
\caption{(a) Benchmark with a normalized $128\times128$ DCT basis matrix. (b) Diagonal-direction frequency map, with dark = low and bright = high diagonal frequency, computed from (a).
(c) Cheng2020-Anchor ($q{=}6$) reconstruction: the low-frequency upper-left region remains close to the input, whereas the high-frequency lower-right region is visibly altered.}
\label{fig:dct_example_rgb}
\end{figure}

\begin{figure*}[t]
\vspace{-1em}
  \centering
  \includegraphics[width=\textwidth]{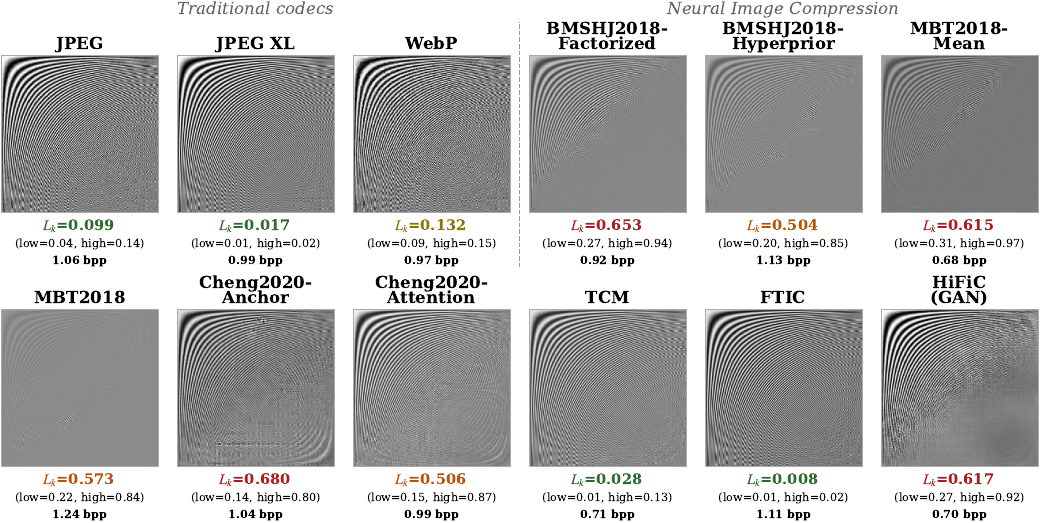}
  \vspace{-1em}
  \caption{DCT-basis response at the nearest available operating point to $\approx$1.0\,bpp (per-panel bpp shown), $256\times256$, for eleven codecs plus the GAN-based HiFiC codec (bottom-right, at its highest fixed rate, $0.70$\,bpp; best viewed zoomed in). The three classical codecs are left of the dashed line; the learned codecs follow. Under each panel: median leakage $L_k$ ($\downarrow$ better, colored green$\to$red by magnitude), its \textit{low}/\textit{high}-band values, and the measured bitrate. At these nearby rates the high-frequency leakage gap between learned and classical codecs persists ($L_{\mathrm{high}}$: $0.02$-$0.15$ for traditional codecs vs.\ $0.80$-$0.97$ for the BMSHJ/MBT/Cheng families); TCM and FTIC fall in the classical range under this diagnostic. HiFiC ($L_k{=}0.62$) shows structured content in the region with high measured leakage rather than the smooth blur of the MSE-trained decoders.}
  \label{fig:ex1_dct_grid_allmodels}
\end{figure*}

\noindent\textit{Controlled spectral probing.}\enspace
We therefore introduce a DCT-basis diagnostic benchmark for measuring frequency-dependent codec output. It uses two probes: a composite image formed from the full DCT basis, which the codec processes jointly as one stress image, and isolated single-frequency inputs that place one component per image. From codec reconstructions we derive a frequency-response matrix and diagnostics for target-frequency gain and the distribution of output energy across frequencies. Only the isolated probes admit an independent per-frequency interpretation, in the spirit of sine-sweep testing but with the caveat that practical codecs are nonlinear and content adaptive; the composite basis is a joint stress test rather than a set of independent measurements.

\begin{figure*}[t]
  \centering
  \begin{minipage}[t]{0.29\linewidth}
    \centering
    \includegraphics[width=\linewidth]{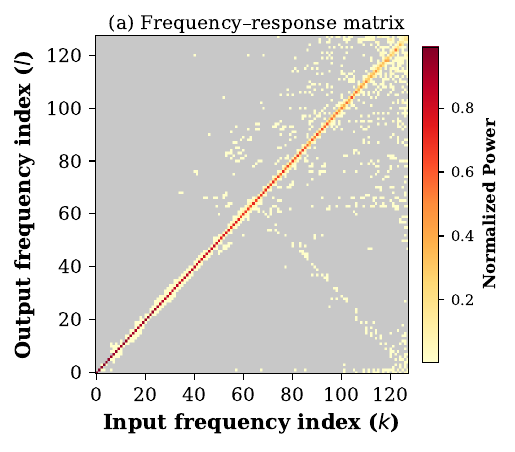}
    \vspace{-1.5em}
  \end{minipage}%
  \hfill
  \begin{minipage}[t]{0.41\linewidth}
    \centering
    \includegraphics[width=\linewidth]{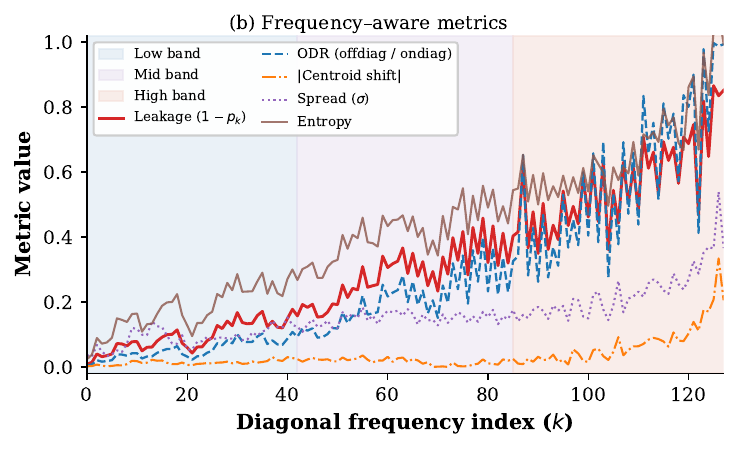}
    \vspace{-1.5em}
  \end{minipage}%
  \hfill
  \begin{minipage}[t]{0.29\linewidth}
    \centering
    \includegraphics[width=\linewidth]{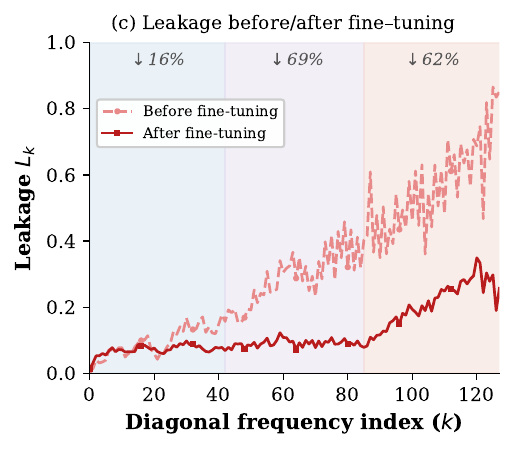}
    \vspace{-1.5em}
  \end{minipage}
  \caption{Frequency-domain analysis of Cheng2020-Anchor on a $128\times128$ DCT basis ($q=6$).
  (a)~The frequency-response matrix shows strong off-target redistribution in high-frequency components (top-right) and near-perfect target-bin purity in low-frequency components (bottom-left).
  (b)~Frequency-aware metrics quantify these effects across low, mid, and high bands (in the panel legend, $p_k$ denotes the retention $R_{k,k}$, so leakage is $1-p_k$).
  (c)~Leakage $L_k$ before vs.\ after decoder-only fine-tuning with the proposed leakage loss ($900$ steps; protocol in Sec.~\ref{new:sec:finetune}); percentages indicate per-band reduction.}
  \label{fig:freq_analysis}
  \end{figure*}

\paragraph{Practical scope}
The controlled-probe evaluation complements natural-image evaluation. Conventional metrics
quantify aggregate distortion; the controlled probes localize changes in the
measured spectral response (Figs.~\ref{fig:ex1_dct_grid_allmodels}
and~\ref{fig:freq_analysis}), while their relation to natural-image error is
summarized only by descriptive alignment statistics (Table~\ref{tab:main_results}).
The observed failures are content-dependent rather than universal. In the
selected Urban100~\cite{huang2015urban100} and DTD~\cite{cimpoi2014dtd} examples, pronounced artifacts occur on repetitive
high-frequency structures such as facade slats, lace, fine line patterns,
zigzags, and meshes, whereas an irregular texture in the same gallery is much
less affected (Suppl.\ Figs.~\ref{sfig:gallery1}-\ref{sfig:gallery2}).
Such localized failures can be diluted by dataset-level averages. Controlled
probes are therefore useful for codec and checkpoint stress testing,
regression testing across operating points, and decoder regularization; they
are not intended to predict the error of an individual natural image.

\paragraph{Contributions}
We make three contributions:
\begin{itemize}[leftmargin=1.4em,nosep]
\item \textbf{Controlled DCT benchmark:} We combine a basis DCT matrix image and
isolated single-frequency probes, with relative off-target energy and gain as
its primary outputs (Fig.~\ref{fig:freq_analysis}); at nearby measured rates,
the older BMSHJ, MBT, and Cheng2020 models have high-band leakage of
$0.80$-$0.97$ and gains of $0.02$-$0.23$, whereas TCM and FTIC fall in the
classical-codec range under this diagnostic
(Fig.~\ref{fig:ex1_dct_grid_allmodels}, Table~\ref{tab:gain_basis},
Fig.~\ref{fig:leakage_vs_bpp_256});
\item \textbf{Error analysis and failure diagnosis:} We derive exact error decompositions and leakage-error bounds
(Eqs.~\eqref{sf:eq:exact}-\eqref{new:eq:frob_upper}) and use isolated probes to identify a frequency-specific instability in the
released Cheng2020-Anchor checkpoints (Fig.~\ref{sf:fig:artifact_visual}); two
tested input-independent response models do not transfer quantitatively to
natural images in the evaluated setting, motivating a descriptive alignment
summary over 115 images (Table~\ref{tab:main_results});
\item \textbf{Decoder-only leakage regularization:} Across all eight evaluated architectures,
fine-tuning with natural-image and leakage losses reduces basis off-target
energy by $90$-$99.8\%$ while keeping the bitstream fixed, at a Kodak PSNR
cost of $0.09$-$0.59$\,dB
(Table~\ref{new:tab:ft_joint}, Fig.~\ref{new:fig:ft_natural}).
\end{itemize}
\section{Related Work}
\label{sec:related}

\paragraph{Transform coding and learned-codec evaluation}
Classical codecs make frequency allocation explicit through transforms and
coefficient quantization, with the DCT underlying JPEG and related coding
standards~\cite{ahmed1974dct,wallace1991jpeg,sullivan2012hevc}, and learned models have been
applied in the DCT domain itself, e.g. for artifact removal~\cite{sun2020dct}.
Learned image
compression instead optimizes nonlinear transforms and entropy models jointly,
progressing from factorized models~\cite{balle2017endtoend} and hyperprior
models~\cite{balle2018hyperprior} to autoregressive priors~\cite{minnen2018joint},
attention mechanisms~\cite{cheng2020attention,mishra2022attention}, space-channel context
models~\cite{he2022elic}, and hybrid transformer-CNN
architectures~\cite{Liu_2023_CVPR}. The approach has been standardized as
JPEG~AI~\cite{alshina2024jpegai}.
Surveys and benchmarks compare these systems primarily through rate-distortion
and perceptual metrics on natural-image datasets
~\cite{ma2019image,Hu2020Learning,Yang2022An}. Such evaluation establishes
overall coding efficiency but does not identify the response to a particular
input frequency. We use the DCT basis as a controlled stimulus rather than as a
codec transform, and retain natural-image metrics as a separate
evaluation axis.

\paragraph{Spectral bias and codec diagnostics}
Spectral-bias studies show that neural networks tend to learn
low-frequency \emph{functions}~\cite{rahaman2019spectral}, not the frequency
response of a rate-constrained codec. Lieberman et al.~\cite{lieberman2023spectral}
used Fourier summaries of reconstruction residuals to study NIC
generalization and robustness. This is an \emph{a posteriori} analysis of
natural or corrupted images, whereas our protocol controls the input
spectrum. The composite basis yields a response matrix, while isolated inputs
provide independent per-frequency measurements
(Sec.~\ref{sf:sec:single_freq}); target gain and off-target energy are measured
separately. Natural-image spectra and controlled probes are complementary;
we do not assume quantitative transfer between them
(Sec.~\ref{sf:sub:natural}). The contribution is the controlled-probe protocol
and its error decomposition.

\paragraph{Frequency-aware codec design}
Frequency-aware codecs use explicit decomposition~\cite{gao2021frequency},
FTIC's directional
attention~\cite{li2024ftic}, WeConvene's wavelet convolution and entropy
model~\cite{fu2024weconvene}, auxiliary wavelet transforms
~\cite{li2025disentangled}, content-adaptive transforms
~\cite{ma2026adaptivefrequency}, or multiscale regularization of energy
dispersion and aliasing~\cite{cong2026spectralregularization}. These methods
modify the architecture or training objective to improve rate-distortion
performance. Our model-agnostic benchmark instead measures target-frequency gain and
off-target redistribution in trained codecs under the same controlled inputs.
The TCM and FTIC results are diagnostic comparisons, not causal evidence about
individual architectural components.

\paragraph{Robustness and distribution shift}
NIC robustness has been studied through common corruptions
~\cite{lieberman2023spectral}, adversarial rate or distortion attacks and
fine-tuning defenses~\cite{Chen2021Toward,liu2024malice,kalmykov2026t},
robustness benchmarks~\cite{bychkov2025nicrobustbench}, and
instability under successive compression~\cite{kim2022successive}. These works
concern distribution shift, repeated coding, or deliberately optimized perturbations. Our
isolated gratings are instead structured stress stimuli with known spectral
support. The observed Cheng2020-Anchor failure is reported only for the released
checkpoint family (Fig.~\ref{sf:fig:artifact_visual}); the experiment neither
isolates an architectural cause nor characterizes behavior on deployment data.

\paragraph{Perceptual and generative compression}
Generative compression methods address the rate-distortion-perception
trade-off~\cite{blau2019rdp}. Examples include HiFiC
~\cite{mentzer2020hific}, implicit local-likelihood models
~\cite{muckley2023illm}, and conditional diffusion decoders
~\cite{yang2023conditionaldiffusion}. Some spectral-consistency objectives use
high-frequency discriminators~\cite{jiang2026sscl}. Perceptual or
distributional fidelity does not directly measure target-frequency
preservation. We include HiFiC as a perceptual reference and report its relative
off-target energy. We do not generalize its measured
behavior to other generative codecs.

\section{Methodology}
\label{sec:methodology}

\subsection{DCT-Basis Probe}
\label{sec:dct_properties}

The orthonormal DCT basis can be assembled columnwise as an $n\times n$
composite image $\mathbf D$ (Fig.~\ref{fig:dct_example_rgb}(a)). Frequencies
increase from the upper-left to the lower-right region. Each column is a
single-frequency pattern, but the codec processes all columns jointly as one
image. Independent per-frequency analysis uses the
isolated probes of Sec.~\ref{sf:sec:single_freq}. Before compression,
\[
\mathbf D^\top\mathbf D=\mathbf I
\]
so each column is localized to one DCT bin. Compression may change the target
coefficient and move output energy to other bins; target-frequency gain and
relative off-target energy quantify these effects separately
(Table~\ref{tab:gain_basis}). Notation is collected in Suppl.\
Table~\ref{tab:supp_notation}.

The assembled basis also forms diagonal chirp trajectories through Nyquist
folding of its quadratic phase (Fig.~\ref{fig:dct_example_rgb}(b)); their
derivation and interpretation are given in Suppl.\ Sec.~\ref{sec:supp_chirp}.

\subsection{Response Matrix and Metrics}
\label{sec:dct_analysis}
\label{sec:metrics}

Let $d_k$ be column $k$ of $\mathbf D$ and $\widehat d_k$ the corresponding
column after joint compression and reconstruction of the composite image. We
apply the 1-D DCT $\mathcal D(\cdot)$ to each reconstructed column and, for
positive output energy, define the \emph{frequency-response matrix}
\begin{equation}
R_{i,k} = \frac{|\mathcal{D}(\widehat{d}_k)_i|^2}
               {\sum_{j=0}^{n-1} |\mathcal{D}(\widehat{d}_k)_j|^2}, \label{eq::response_matrix}
\end{equation}
where $R\in\mathbb R^{n\times n}$. Each column of $R$ sums to one:
$R_{k,k}$ is the output-energy fraction in the target bin, and $R_{i,k}$ for
$i\ne k$ is the fraction assigned to bin $i$. Lossless reconstruction gives
$R=I$. Equation~\eqref{eq::response_matrix} applies to positive-energy columns,
as in all evaluated runs. The implementation uses $\epsilon=10^{-12}$ in
numerical denominators and assigns $R_{:,k}=0$ and $L_k=1$ to a zero-output
column; exact identities in Suppl.\ Sec.~\ref{sec:supp_metric_bounds} use the
positive-energy definition above.

The columnwise transform preserves the index $k$ of each target pattern. A 2-D
DCT would instead describe the joint spatial layout of the composite image and
would not yield the same column-indexed response.

We use two primary diagnostics, $L_k$ and $g_k$; auxiliary quantities describe
the response shape.

\begin{itemize}[leftmargin=1em]
  \item \textbf{Relative off-target energy (leakage):} Fraction of the
  reconstructed column's energy outside the target bin,
\begin{align}
  L_k = 1 - R_{k,k} \in [0,1],
   \label{eq:leakage}
\end{align}
  which measures normalized redistribution and is unchanged by uniform
  amplitude scaling.

  \item \textbf{Target-frequency gain:} Signed target-bin coefficient,
  $g_k=\mathcal D(\widehat d_k)_k$, for a unit input coefficient. It measures
  target amplitude. If $\widehat d_k=\alpha d_k$, then $L_k=0$ but
  $g_k=\alpha$; reporting both separates attenuation from redistribution.

  \item \textbf{Off-Diagonal Ratio (ODR):} Bounded ratio of off-target to
  target energy,
\[
  \mathrm{ODR}^{\mathrm{norm}}_k = \tanh\!\left[\frac{\sum_{i\neq k} R_{i,k}}{2(R_{k,k}+\epsilon)}\right],
\]
  used only as a display-scale reparameterization of $L_k$.

  \item \textbf{Centroid Shift:} Normalized displacement of the output-energy
  centroid from the target index,
\[
  \mu_k = \sum_i i\,R_{i,k}, \quad
  \Delta_k = \frac{|\mu_k - k|}{\max(k,\;n{-}1{-}k)}.
\]

  \item \textbf{Spread:} Standard deviation about $\mu_k$, normalized by the
  distance to the farther band edge,
\[
  \sigma^{\mathrm{norm}}_k =
  \frac{\sqrt{\sum_i (i-\mu_k)^2 R_{i,k}}}
       {\max(\mu_k,\;n{-}1{-}\mu_k)+\epsilon}.
\]
  \item \textbf{Entropy:} Normalized Shannon entropy of the output-energy
  distribution,
\[
  H^{\mathrm{norm}}_k = \frac{-1}{\log n} \sum_i R_{i,k}\log(R_{i,k}+\epsilon),
  \]
\end{itemize}
Low $\Delta_k$, $\sigma^{\mathrm{norm}}_k$, and $H^{\mathrm{norm}}_k$ indicate
limited displacement, dispersion, and spectral disorder, respectively. Pooled
Spearman correlations with $L_k$ are $0.98$ for entropy and $0.92$ for
spread. Quantitative comparisons therefore
use $L_k$ and $g_k$, while the auxiliary metrics interpret Fig.~\ref{fig:freq_analysis}(b).
Formal identities and pooled analyses are given in Suppl.\
Sec.~\ref{sec:supp_metric_bounds}.

\subsection{Isolated Single-Frequency Probes}
\label{sf:sec:single_freq}
The basis-matrix probe processes all $n$ basis patterns jointly. For independent
per-frequency measurements, we instead place one 2-D DCT component in each
test image. Let $\mathbf d_k\in\mathbb R^n$ be column $k$ of $\mathbf D$ and
$\mathbf 1\in\mathbb R^n$ the all-ones vector. We define two unit-energy,
rank-1 probes:
\begin{align}
  \mathbf{X}_k^{(1)} &= n^{-1/2}\,\mathbf{d}_k \mathbf{1}^\top
    \;\in\mathbb{R}^{n\times n},
    \label{sf:eq:X1}\\
  \mathbf{X}_k^{(2)} &= \mathbf{d}_k \mathbf{d}_k^\top
    \;\in\mathbb{R}^{n\times n},
    \label{sf:eq:X2}
\end{align}
with $\|\mathbf{X}_k^{(1)}\|_F=\|\mathbf{X}_k^{(2)}\|_F=1$; both are shown in Fig.~\ref{sf:fig:probe_construction}.

For codec input, each probe is embedded into $[0,1]$ by the centered affine map
$x\mapsto\tfrac12+s x$, using one fixed scale $s$ for all $k$, replicated over
three channels, and mapped back after decoding. We do not use min-max
normalization because it introduces a $k$-dependent DC offset and changes the
measured leakage.

\smallskip\noindent\textbf{Single-bin spectra.}
Let $\mathcal{D}_2(\mathbf{X})=\mathbf{D}^\top\mathbf{X}\mathbf{D}$ denote
the 2-D DCT. The probes satisfy
\[
  \mathcal D_2(\mathbf X_k^{(1)})=\mathbf e_k\mathbf e_0^\top,
  \qquad
  \mathcal D_2(\mathbf X_k^{(2)})=\mathbf e_k\mathbf e_k^\top.
\]
Thus $\mathbf X_k^{(1)}$ and $\mathbf X_k^{(2)}$ occupy the single unit-amplitude
bins $(k,0)$ and $(k,k)$, respectively. The proof and a controlled attenuation
check are given in Suppl.\ Sec.~\ref{sec:supp_singlefreq}.

$\mathbf X_k^{(1)}$ varies along one spatial axis, whereas
$\mathbf X_k^{(2)}$ varies along both. Their responses are analyzed separately;
their error ratio is not used as an orientation measure. After inverse affine
mapping, reconstructed energy outside the target bin is attributed to the
codec response.

\subsection{Exact Error Decomposition and Leakage Bounds}
\label{sf:sub:distortion}

Let $\mathbf X$ be either probe, with unit coefficient in target bin $(a,b)$,
and let $\widehat{\mathbf G}=\mathcal D_2(\widehat{\mathbf X})$ be the 2-D DCT
of its reconstruction. Define the \emph{target-frequency gain}
$g_k=\widehat G_{a,b}$ and \emph{off-bin energy}
$E^{\mathrm{off}}_k=\sum_{(i,j)\neq(a,b)}\widehat{G}_{i,j}^2$. By Parseval's identity, every reconstruction satisfies

\begin{equation}
  \label{sf:eq:exact}
  \|\mathbf{X}-\widehat{\mathbf{X}}\|_F^2
  \;=\; (1-g_k)^2 \;+\; E^{\mathrm{off}}_k .
\end{equation}

The two terms are the target-coefficient error and the energy outside the
target bin.

Define the single-frequency \emph{energy leakage}
$\lambda_k$ for $\widehat{\mathbf G}\neq0$ as
\[
\lambda_k
=\frac{E_k^{\mathrm{off}}}{\|\widehat{\mathbf G}\|_F^2}
=1-\frac{g_k^2}{\|\widehat{\mathbf G}\|_F^2}.
\]
This is the isolated-probe analogue of $L_k$ in Eq.~\eqref{eq:leakage}; for an
all-zero reconstruction, we set $\lambda_k=1$. With this convention, every
reconstruction satisfies
\begin{equation}
  \label{sf:eq:bounds_lower}
  \lambda_k \;\le\; \|\mathbf{X}-\widehat{\mathbf{X}}\|_F^2.
\end{equation}
If $\|\widehat{\mathbf G}\|_F=1$ and $g_k\ge0$, then
\begin{equation}
  \label{sf:eq:bounds_upper}
  \|\mathbf{X}-\widehat{\mathbf{X}}\|_F^2 \;\le\; 2\lambda_k .
\end{equation}
The lower bound requires no output-energy normalization; the upper bound does.
Without this constraint, no finite upper bound can depend on $\lambda_k$ alone.
Proofs, equality conditions, and optimality of the constants are given in
Suppl.\ Sec.~\ref{sec:supp_metric_bounds}.

\subsection{Full-Basis Error Bounds}
\label{new:sec:theory}
Exact relations among the auxiliary response metrics are given in Suppl.\
Sec.~\ref{sec:supp_metric_bounds}. Here we apply the single-frequency bounds
columnwise to the basis-matrix probe. Let
$\widehat{\mathbf D}=[\widehat{\mathbf d}_0,\ldots,\widehat{\mathbf d}_{n-1}]$
be its reconstruction and $\bar L=n^{-1}\sum_k L_k$ the mean leakage. If every
reconstructed column has positive energy, then
\begin{equation}
  \label{new:eq:frob_lower}
  \tfrac{1}{n}\,\|\widehat{\mathbf{D}}-\mathbf{D}\|_F^2 \;\ge\; \bar L ,
\end{equation}
without assuming energy preservation. If, in addition, each column satisfies
$\|\mathcal D(\widehat{\mathbf d}_k)\|_2=1$ and
$\mathcal D(\widehat{\mathbf d}_k)_k\ge0$, then
\begin{equation}
  \label{new:eq:frob_upper}
  \tfrac{1}{n}\,\|\widehat{\mathbf{D}}-\mathbf{D}\|_F^2 \;\le\; 2\bar L .
\end{equation}
Thus Eq.~\eqref{new:eq:frob_lower} bounds mean per-column squared error without
output-energy preservation, whereas Eq.~\eqref{new:eq:frob_upper} requires the
stated normalization and sign conditions. Proofs are given in Suppl.\
Sec.~\ref{sec:supp_metric_bounds}.

Weighting the profiles by a contrast-sensitivity function leaves the codec
ranking essentially unchanged ($\rho_s{=}0.991$; Suppl.\
Sec.~\ref{new:sub:csf_weight}).

\subsection{Natural-Image Spectral Alignment}
\label{sec:weighted_leakage}

Leakage $L_k$~\eqref{eq:leakage} is measured on synthetic inputs. To compare
its frequency profile with natural-image distortion, we map both to a common
radial-frequency axis. For image $X$, reconstruction $\widehat X$, and error
$e=X-\widehat X$, let $P_X(f)$ and $P_e(f)$ be their 2-D DCT powers averaged
within $N_b$ radial rings. Define the \emph{radial distortion fraction}
\begin{equation}
\rho(f) = \frac{P_e(f)}{P_X(f)+P_e(f)+\epsilon},
\qquad \rho(f)\in[0,1].
\end{equation}

We linearly interpolate $L_k$ at
$k(f)=f(n{-}1)/f_{\max}$, where $f_{\max}$ is the largest radial ring
frequency, clamped to $[0,n{-}1]$, to obtain $L^{\mathrm r}(f)$.
This mapping assumes isotropy and serves only as a radial proxy. We define
\begin{equation}
\tilde{\mathcal{L}}(X,\widehat{X}) = \frac{1}{N_b}\sum_{f=0}^{N_b-1} \rho(f)\,L^{\mathrm{r}}(f).
\end{equation}
For $\bar\rho=N_b^{-1}\sum_f\rho(f)>0$, the ratio
$\tilde{\mathcal L}/\bar\rho$ is the $\rho$-weighted mean of
$L^{\mathrm r}(f)$: it is high when measured distortion concentrates at
frequencies where the mapped synthetic profile leaks, and low when the two
profiles do not overlap. It summarizes spectral alignment only; it is neither a
distortion predictor nor a causal attribution. Transfer results and limitations
are reported in Sec.~\ref{sec:distortion_decomp}.

\subsection{Leakage-Regularized Fine-Tuning Loss}
\label{sec:metric_loss}

We fine-tune the decoder using two loss terms. The natural-image term is the
mean pixel MSE over a batch $\mathcal B$ of image patches,
\begin{subequations}
\begin{equation}
\mathcal L_{\mathrm{dist}}
=\frac{1}{|\mathcal B|}\sum_{\mathbf X\in\mathcal B}
\operatorname{MSE}(\widehat{\mathbf X},\mathbf X).
\label{eq:dist_loss}
\end{equation}
The frequency term is computed on a fixed composite DCT-basis probe. For its
reconstruction $\widehat{\mathbf D}$, we compute $R$ from
Eq.~\eqref{eq::response_matrix}, set $L_k=1-R_{k,k}$, and define
\begin{equation}
    \mathcal{L}_{\mathrm{freq}} = \sum_{k=0}^{n-1} w_k L_k.
    \label{eq::loss_function}
\end{equation}
\end{subequations}
We set $k_0=\lfloor0.33n\rfloor$, so this term acts on approximately the upper
two-thirds of the frequency indices. Within the active band, the weights
increase with frequency:
\begin{equation}
  \widetilde w_k=
  \begin{cases}
    0, & k<k_0,\\
    \bigl(k/(n-1)\bigr)^p+0.2, & k\ge k_0,
  \end{cases}
  \qquad
  w_k=\frac{\widetilde w_k}{\sum_j\widetilde w_j}.
\end{equation}
The constant $0.2$ provides a fixed baseline weight within the active band.
We use $p{=}3$ for leakage-regularized fine-tuning.

Because $\mathcal L_{\mathrm{freq}}$ targets only $k\ge k_0$, two additional
terms constrain the low-frequency response. Let $R^{(0)}$ denote the response
matrix before fine-tuning, and let $\mathbf D_{<k_0}$ and
$\widehat{\mathbf D}_{<k_0}$ denote the target and reconstructed submatrices
formed by the first $k_0$ columns, respectively:
\begin{align}
\mathcal{G}_{\mathrm{pur}}
&=\frac{1}{k_0}\sum_{k<k_0}
  \max\bigl(0,R^{(0)}_{k,k}-R_{k,k}-0.02\bigr), \\
\mathcal{G}_{\mathrm{mse}}
&=\operatorname{MSE}\!\left(\widehat{\mathbf D}_{<k_0},
\mathbf D_{<k_0}\right).
\end{align}
$\mathcal G_{\mathrm{pur}}$ penalizes low-band purity decreases beyond a fixed
$0.02$ margin, and $\mathcal G_{\mathrm{mse}}$ penalizes low-band
reconstruction error. The full objective is
\begin{equation}
\mathcal{L}=\mathcal{L}_{\mathrm{dist}}
 +\lambda_{\mathrm{freq}}\mathcal{L}_{\mathrm{freq}}
 +0.12\mathcal{G}_{\mathrm{pur}}+0.02\mathcal{G}_{\mathrm{mse}}.
\label{eq:joint_objective}
\end{equation}
The MSE-only control uses only $\mathcal L_{\mathrm{dist}}$ in
Eq.~\eqref{eq:dist_loss}.
Leakage-regularized fine-tuning uses the full objective in
Eq.~\eqref{eq:joint_objective}. We select
$\lambda_{\mathrm{freq}}\in\{0.003,0.01,0.03\}$ using the held-out DIV2K
monitor (Sec.~\ref{new:sub:ft_joint}; Table~\ref{new:tab:ft_joint}). In both protocols, only the
decoder is updated; the encoder, hyperprior, and entropy model remain fixed, so
the bitstream is unchanged.

\section{Experimental Protocol}
\label{sec:experimental_protocol}

We evaluate frequency-aware rate-distortion performance using a unified pipeline across classical codecs (JPEG, JPEG~XL, WebP) and learned codecs: six models from CompressAI~\cite{begaint2020compressai} (BMSHJ2018-Factorized, BMSHJ2018-Hyperprior~\cite{balle2018hyperprior}, MBT2018-Mean, MBT2018~\cite{minnen2018joint}, Cheng2020-Anchor, Cheng2020-Attention~\cite{cheng2020attention}), as well as TCM~\cite{Liu_2023_CVPR} and FTIC~\cite{li2024ftic}. As a perceptual anchor we additionally profile the GAN-based HiFiC~\cite{mentzer2020hific} at its three released operating points (Figs.~\ref{fig:ex1_dct_grid_allmodels} and~\ref{fig:leakage_vs_bpp_256}); its rates are fixed by the released checkpoints, so it enters the anchored comparisons rather than the nearest-rate comparisons.
For each tested image size ($n \in \{64,128,256,512,1024\}$), we generate normalized three-channel DCT basis images and compress them at multiple quality levels or bitrates.
The decompressed outputs are mapped back to the basis scale and analyzed to obtain the frequency-response matrix $R$ (Section~\ref{sec:dct_analysis}) and the proposed metrics. The basis figures use each codec's available setting nearest $1.0$\,bpp ($0.68$-$1.24$\,bpp; HiFiC is fixed at $0.70$\,bpp; Fig.~\ref{fig:ex1_dct_grid_allmodels}). For natural images, per-dataset settings yield codec-level mean rates of $0.82$-$0.98$\,bpp (Table~\ref{tab:main_results}). All stimuli are achromatic (the basis replicated across RGB); chromatic probing, relevant given the visibly chromatic artifacts of Figs.~\ref{fig:dct_example_rgb}(c) and~\ref{fig:dct_leakage}(b) and the chroma subsampling of classical codecs, is left to future work. Full reproducibility details are given in Suppl.\ Sec.~\ref{sec:supp_repro}.

\FloatBarrier
\section{Results and Discussion}
\label{sec:results}

\subsection{Spectral Allocation Across Codecs}
\label{sec::example1}
\begin{figure}[t]
  \centering
  \includegraphics[width=\columnwidth]{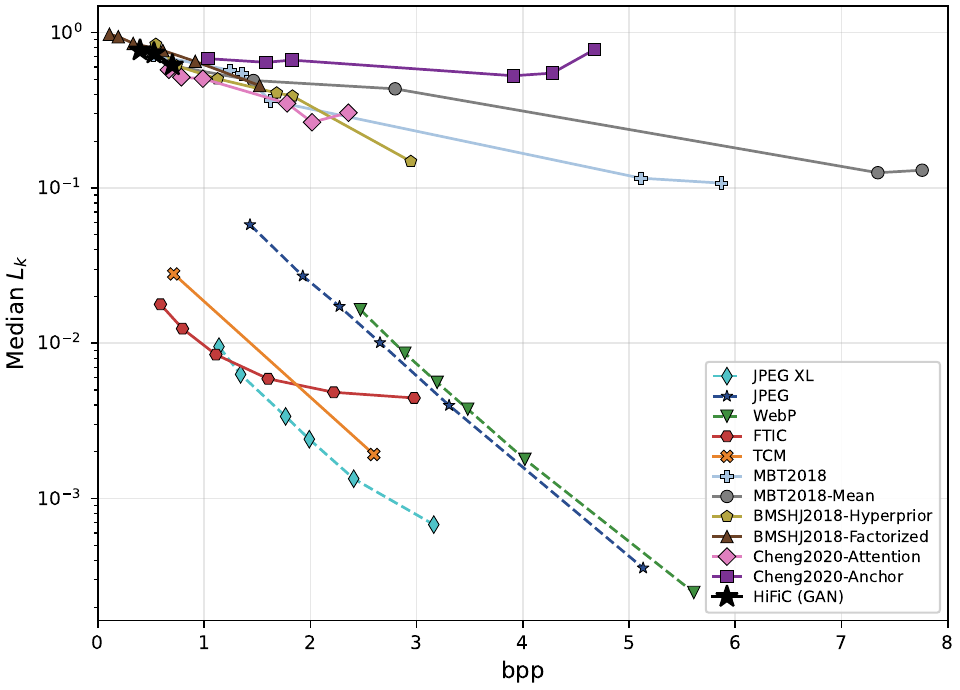}
  \vspace{-1.5em}
  \caption{DCT rate-leakage comparison at $256 \times 256$. Traditional codecs (dashed) achieve lower median leakage across all bitrates than the older BMSHJ/MBT/Cheng models (solid), while TCM and FTIC approach classical levels. The GAN-based HiFiC is shown at its three fixed operating points (black stars).}
  \label{fig:leakage_vs_bpp_256}
\end{figure}

Figure~\ref{fig:ex1_dct_grid_allmodels} compares eleven codecs at their available settings nearest $1.0$\,bpp and HiFiC at its highest fixed rate. The separation in median leakage persists across the evaluated rate sweep (Fig.~\ref{fig:leakage_vs_bpp_256}).

JPEG, JPEG~XL, and WebP have high-band relative off-target energy $L_{\mathrm{high}}{\leq}0.15$. The BMSHJ/MBT/Cheng families reach $L_{\mathrm{high}}{=}0.80$-$0.97$ and differ mainly in their low- and mid-band response. TCM and FTIC have overall median leakage $L_k{\leq}0.03$ and fall in the classical-codec range. Cheng2020-Anchor has the highest median leakage: $L_k{=}0.68$ at $256{\times}256$ ($q{=}1$; Fig.~\ref{fig:ex1_dct_grid_allmodels}) and $0.73$ at $512{\times}512$ ($q{=}6$; Table~\ref{tab:main_results}).

Table~\ref{tab:gain_basis} separates target-frequency gain from off-target redistribution. The BMSHJ, MBT, and Cheng2020 models have high-band gains of $0.02$-$0.23$, compared with $0.78$ and $0.97$ for TCM and FTIC. Thus, high off-target energy co-occurs with low target gain at these settings, but the two quantities remain distinct.

\begin{table}[t]
\centering
\caption{Composite-basis response of learned codecs at their settings nearest
$1.0$\,bpp ($256\times256$). Values are means over the top-quarter band
($k\ge3n/4$); best values are bold. Classical codecs are not included.}
\label{tab:gain_basis}
\setlength{\tabcolsep}{5pt}
\small
\begin{tabular}{@{}lccc@{}}
\toprule
Codec & bpp & $L_{\mathrm{high}}$ $\downarrow$ & gain $g_k$ $\uparrow$ \\
\midrule
BMSHJ2018-Factorized & $0.92$ & $0.94$ & $0.05$ \\
BMSHJ2018-Hyperprior & $1.13$ & $0.85$ & $0.09$ \\
MBT2018-Mean         & $0.68$ & $0.97$ & $0.02$ \\
MBT2018              & $1.24$ & $0.84$ & $0.23$ \\
Cheng2020-Anchor     & $1.04$ & $0.80$ & $0.20$ \\
Cheng2020-Attention  & $0.99$ & $0.87$ & $0.15$ \\
\midrule
TCM                  & $0.71$ & $0.13$ & $0.78$ \\
FTIC                 & $1.11$ & $\mathbf{0.02}$ & $\mathbf{0.97}$ \\
\bottomrule
\end{tabular}

\end{table}
Rate selection on the basis yields harsher classical-codec settings than rate selection on natural images. Figure~\ref{fig:ex1_dct_grid_allmodels} and Table~\ref{tab:main_results} therefore state their protocols separately; their classical-codec leakage values are not directly interchangeable. TCM uses $\lambda{=}0.05$ ($p{=}128$) in Table~\ref{tab:main_results} and its $p{=}64$ variant in the basis-rate comparisons.

\subsection{Single-Frequency Response}
\label{sf:sub:consistency}

Composite-basis leakage $L_k$ measures columns processed jointly, whereas
$\lambda_k$ is obtained from isolated inputs. With decoder outputs clamped to
$[0,1]$ at settings nearest $1$\,bpp, the profiles agree on the
$\mathbf{X}^{(2)}$ sweep: Spearman's $\rho_s$ is $0.90$-$0.97$ for the eight
learned codecs and $0.89$ for JPEG. Correlation is lower for JPEG~XL
($\rho_s{=}0.37$), where leakage is near the numerical floor (median
$\lambda_k{\approx}6{\cdot}10^{-3}$), and for WebP on the
$\mathbf{X}^{(1)}$ gratings ($\rho_s{=}0.59$), where the low-contrast
gratings are strongly suppressed at this setting (median
$\lambda_k{\approx}0.6$). For Cheng2020-Anchor at its highest quality, removing the clamp reduces
the correlation to approximately zero because extreme outputs dominate the
profile (Sec.~\ref{sf:sub:instability}).

The two constructions $\mathbf{X}^{(1)}{=}(k,0)$ and
$\mathbf{X}^{(2)}{=}(k,k)$ differ in contrast, radial frequency, and spatial
structure. Their responses are therefore reported as complementary probe
geometries and are not used to claim an isolated orientation effect.

\begin{figure*}[t]
  \centering
  \includegraphics[width=0.84\textwidth]{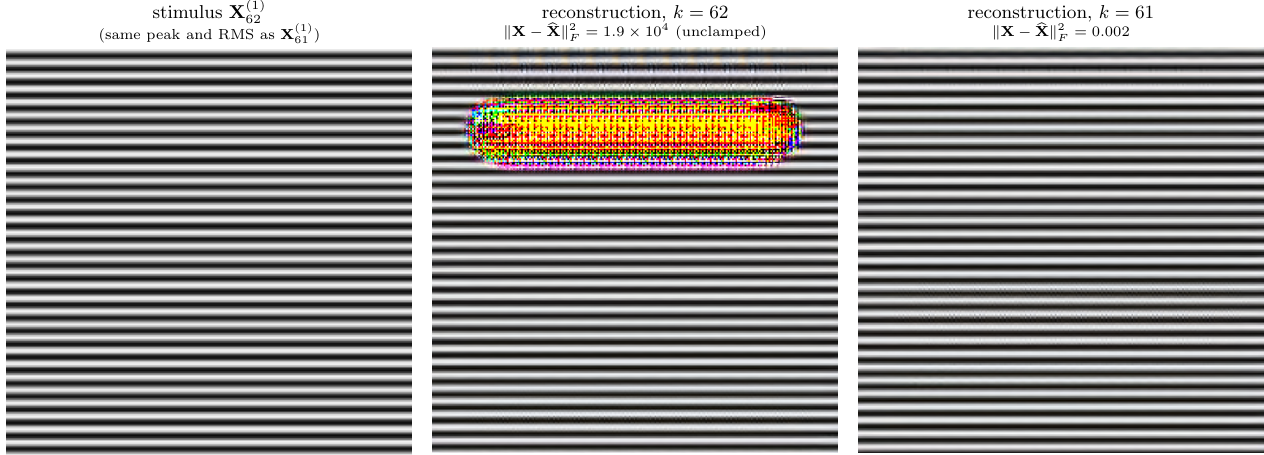}
  \caption{Released-checkpoint instability of Cheng2020-Anchor: clamped
  reconstructions of neighboring $\mathbf{X}_k^{(1)}$ gratings with matched
  peak and RMS ($n{=}256$, peak $0.45$, $q{=}6$). The $k{=}62$ input
  produces a localized chromatic artifact (pre-clamp error
  $1.9{\times}10^4$), whereas $k{=}61$ is reconstructed with error $0.002$.
  Full unclamped sweeps: Suppl.\ Figs.~\ref{sf:fig:instability}
  and~\ref{sfig:x1gallery}.}
  \label{sf:fig:artifact_visual}
\end{figure*}

\begin{figure}[t]
  \centering
  \begin{subfigure}[t]{0.48\columnwidth}
    \centering
    \vspace{0pt}
    \includegraphics[height=0.20\paperwidth,keepaspectratio]{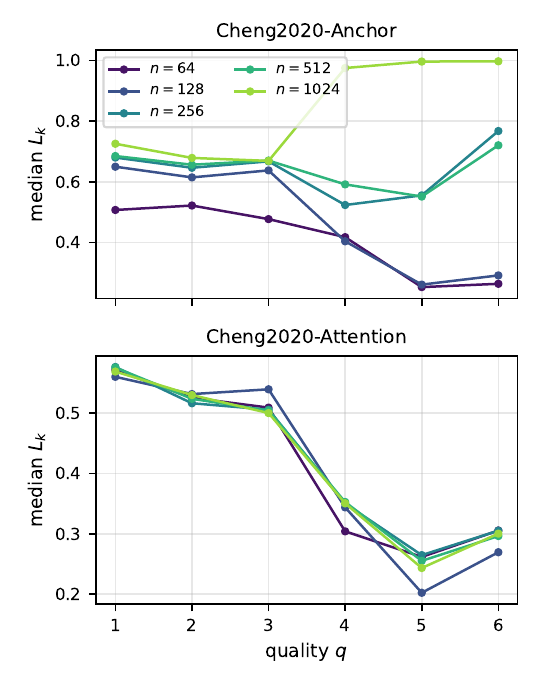}
    \subcaption{Median leakage vs.\ quality $q$ across DCT image sizes
    ($\downarrow$\,better).}
    \label{fig:leakage_vs_quality}
  \end{subfigure}\hfill
  \begin{subfigure}[t]{0.48\columnwidth}
    \centering
    \vspace{0pt}
    \includegraphics[height=0.20\paperwidth,keepaspectratio,trim=0cm 0cm 0cm 0cm,clip=true]{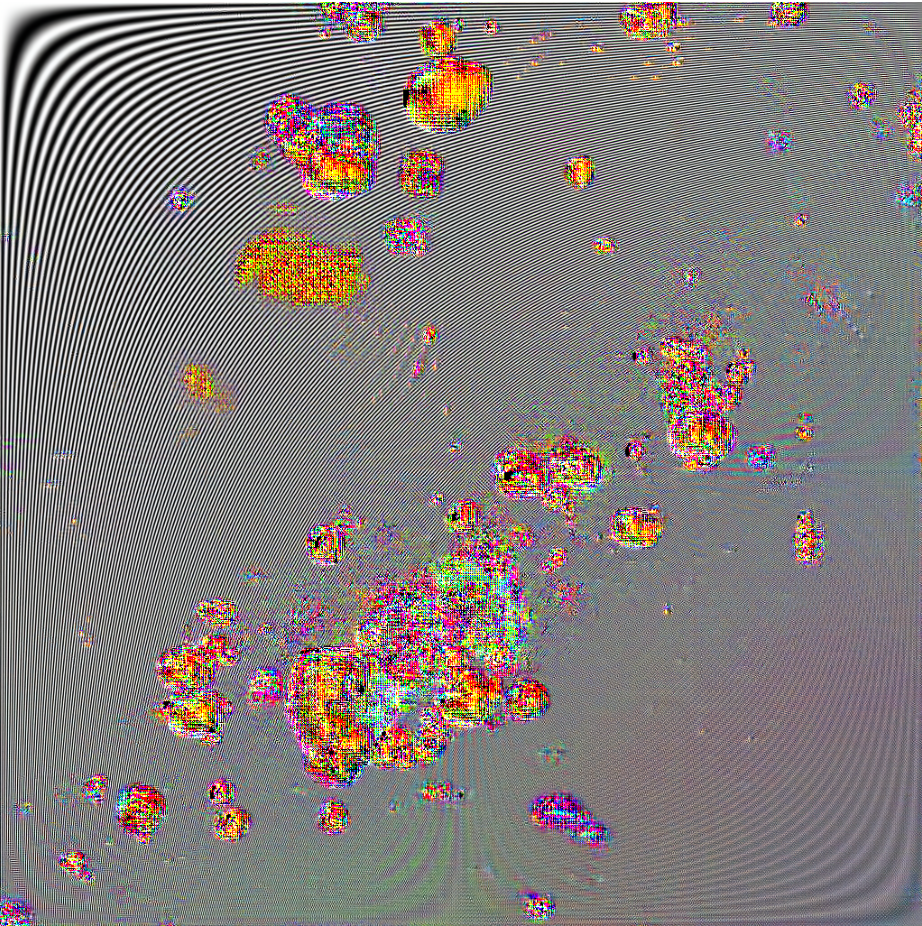}
    \subcaption{Cheng2020-Anchor reconstruction at $1024{\times}1024$
    ($q{=}6$).}
    \label{fig:dct_anchor_1024}
  \end{subfigure}
  \caption{Size and quality dependence of the composite-basis response.
  Cheng2020-Anchor becomes non-monotonic for $n{\geq}256$ and produces
  visible artifacts at $1024{\times}1024$.}
  \label{fig:dct_leakage}
\end{figure}

\subsection{Instability in Cheng2020-Anchor}
\label{sf:sub:instability}

On the $\mathbf{X}_k^{(2)}$ sweep at $n{=}256$, we report the unclamped
per-probe error $\|\mathbf{X}-\widehat{\mathbf{X}}\|_F^2$ of
Eq.~\eqref{sf:eq:exact}; inputs have unit energy. The released
Cheng2020-Anchor checkpoints diverge at isolated frequencies
(Suppl.\ Fig.~\ref{sf:fig:instability}): the $N{=}128$, $q{=}1$-$3$
checkpoints stay below an error of $500$, whereas the $N{=}192$,
$q{=}4$-$6$ checkpoints exceed an error of $10$ on $42$, $62$, and $71$ of
the $256$ frequencies, respectively, reaching $1.9{\times}10^{11}$. Because decoder
width, bitrate, and training instance change together, we restrict this
finding to the released $N{=}192$, $q{=}4$-$6$ checkpoint family;
architecture and training effects are not identified separately. The
$\mathbf{X}_k^{(1)}$ geometry produces the same pattern over broader
high-frequency bands (Suppl.\ Fig.~\ref{sfig:x1gallery}).
Clamping outputs to $[0,1]$ bounds the numerical values but does not repair
the response: the median per-probe error at $q{=}6$ is $0.65$, versus
$0.05$ for the closely related Cheng2020-Attention under the same protocol
(Fig.~\ref{sf:fig:artifact_visual}). On the basis-matrix
probe, Cheng2020-Anchor also becomes non-monotonic for $n{\geq}256$ and
produces visible artifacts at $1024{\times}1024$ (Fig.~\ref{fig:dct_leakage}). Similar
extreme outputs have been reported for these checkpoints under adversarial
perturbations~\cite{kalmykov2026t}; here they are elicited by structured,
low-contrast probes.

\subsection{Transfer to Natural Images}
\label{sec:distortion_decomp}
\label{sf:sub:natural}

The controlled probes characterize each codec's spectral response; here we
delimit how far these measurements alone extend to natural images, using
six Kodak center crops ($512{\times}512$) and four codecs that span the
observed leakage range (Cheng2020-Anchor, Cheng2020-Attention, and
BMSHJ2018-Factorized at $q{=}6$; JPEG at quality $85$).

The prediction target is exact: let $\mathbf{C}=\mathbf{D}^\top\mathbf{X}$ and
$\widehat{\mathbf{C}}=\mathbf{D}^\top\widehat{\mathbf{X}}$. Orthonormality
of $\mathbf D$ gives
\begin{equation}
  \label{sf:eq:exact_decomp}
  \|\mathbf{X}-\widehat{\mathbf{X}}\|_F^2
  = \sum_{k=0}^{n-1}
    \underbrace{\|\mathbf{c}_k-\widehat{\mathbf{c}}_k\|^2}_{\delta_k},
\end{equation}
where $\mathbf{c}_k$ denotes row $k$ of $\mathbf{C}$ and $\delta_k$ is the
exact squared-error contribution of frequency $k$.

Two \emph{input-independent} models, measured once on the basis, predict
$\delta_k$ without seeing the image: pure attenuation,
$\widehat{\mathbf{c}}_k=(1-\alpha_k)\mathbf{c}_k$ with
$\alpha_k{=}1-\mathcal{D}(\widehat{\mathbf{d}}_k)_k$, and the full linear
response $T_{ik}=\mathcal{D}(\widehat{\mathbf{d}}_k)_i$ applied to
$\mathbf{C}$. Neither transfers: the attenuation model overpredicts the
total squared error by factors of $1.7$-$10$ for the learned codecs and
underpredicts JPEG by more than an order of magnitude, and the full
operator overpredicts by factors of $8$-$44$. This conclusion is limited to
the two models tested here.

The mechanism is content adaptivity: rate and quantization depend on the
whole spectrum, so an isolated unit-energy frequency is coded in a
different regime than the same frequency embedded in a natural image, and
no fixed response measured on one stimulus can represent both. What
survives at the codec level is reported in Table~\ref{tab:main_results} as
the descriptive statistics $\tilde{\mathcal{L}}$ and
$\tilde{\mathcal{L}}/\bar{\rho}$ (Sec.~\ref{sec:weighted_leakage}):
computed purely from spectral overlap, they recover the quality ordering of
the eight NIC models ($\rho_s{=}{-}0.98$ and $-0.93$ with PSNR), with the
$\tilde{\mathcal{L}}$ ordering identical on all three datasets. The
added-distortion control (Suppl.\ Sec.~\ref{sec:supp_distcons}) reaches
similar correlations without the leakage weighting.

HiFiC illustrates the distinction between perceptual quality and
target-frequency fidelity: at $0.47$\,bpp on natural images
(basis-probe rate $0.70$\,bpp; Fig.~\ref{fig:ex1_dct_grid_allmodels}), roughly half the rate of the
Table~\ref{tab:main_results} codecs, it attains a lower LPIPS than any of
them ($0.040$), yet the highest normalized alignment
($\tilde{\mathcal{L}}/\bar{\rho}{=}0.78$) and basis leakage of $0.62$-$0.77$
(Fig.~\ref{fig:leakage_vs_bpp_256}).

\begin{table*}[!t]
  \centering
  \caption{Natural-image evaluation of eleven codecs on \textbf{115 images}
  (24 Kodak, 41 CLIC, 50 DIV2K) at per-dataset settings near $0.9$\,bpp.
  Learned codecs use fixed settings; classical codecs are rate-selected per
  dataset. $L_k$ is median leakage on the synthetic $512{\times}512$ basis
  (for classical codecs, the median over three selected settings).
  $\tilde{\mathcal{L}}$ and $\tilde{\mathcal{L}}/\bar{\rho}$ are descriptive
  spectral-alignment statistics. Entries with $\pm$ report mean and standard
  deviation over images; best displayed values in directed columns are bold. Rows are ordered by decreasing $\tilde{\mathcal{L}}$.}
  \label{tab:main_results}
  \renewcommand{\arraystretch}{1.05}
  \setlength{\tabcolsep}{4.5pt}
  \small
  \resizebox{\linewidth}{!}{\begin{tabular}{@{}lccccccc@{}}
\toprule
Codec & $L_k\downarrow$ & bpp & PSNR$\uparrow$ (dB) & MS-SSIM$\uparrow$ & LPIPS$\downarrow$ & $\tilde{\mathcal{L}}$ & $\tilde{\mathcal{L}}/\bar{\rho}$ \\
\midrule
BMSHJ2018-Factorized & $0.455$ & $0.98 \pm 0.40$ & $35.17 \pm 2.77$ & $0.989 \pm 0.005$ & $0.087 \pm 0.064$ & $0.1486 \pm 0.0225$ & $0.6872$ \\
Cheng2020-Anchor & $0.729$ & $0.82 \pm 0.44$ & $36.47 \pm 2.43$ & $0.990 \pm 0.004$ & $0.074 \pm 0.061$ & $0.1372 \pm 0.0402$ & $0.7252$ \\
Cheng2020-Attention & $0.296$ & $0.82 \pm 0.44$ & $36.37 \pm 2.40$ & $0.990 \pm 0.004$ & $0.077 \pm 0.063$ & $0.1067 \pm 0.0192$ & $0.5576$ \\
BMSHJ2018-Hyperprior & $0.146$ & $0.96 \pm 0.48$ & $36.53 \pm 2.03$ & $0.990 \pm 0.004$ & $0.082 \pm 0.063$ & $0.0825 \pm 0.0192$ & $0.4432$ \\
MBT2018-Mean & $0.116$ & $0.95 \pm 0.50$ & $36.71 \pm 1.97$ & $0.990 \pm 0.004$ & $0.079 \pm 0.062$ & $0.0509 \pm 0.0099$ & $0.2799$ \\
MBT2018 & $0.115$ & $0.92 \pm 0.50$ & $36.92 \pm 2.02$ & $0.991 \pm 0.004$ & $0.070 \pm 0.056$ & $0.0453 \pm 0.0088$ & $0.2527$ \\
FTIC & $0.005$ & $0.89 \pm 0.52$ & $37.78 \pm 1.90$ & $\mathbf{0.992 \pm 0.004}$ & $0.064 \pm 0.057$ & $0.0055 \pm 0.0013$ & $0.0347$ \\
TCM & $\mathbf{0.002}$ & $0.91 \pm 0.52$ & $\mathbf{37.95 \pm 1.87}$ & $\mathbf{0.992 \pm 0.004}$ & $\mathbf{0.063 \pm 0.056}$ & $0.0011 \pm 0.0002$ & $0.0069$ \\
\midrule
JPEG & $0.022$ & $0.83 \pm 0.31$ & $31.39 \pm 3.22$ & $0.975 \pm 0.009$ & $0.094 \pm 0.049$ & $0.0100 \pm 0.0011$ & $0.0304$ \\
JPEG XL & $0.004$ & $0.87 \pm 0.32$ & $33.70 \pm 3.13$ & $0.983 \pm 0.010$ & $0.080 \pm 0.037$ & $0.0019 \pm 0.0004$ & $0.0077$ \\
WebP & $0.003$ & $0.92 \pm 0.50$ & $34.08 \pm 2.56$ & $0.982 \pm 0.008$ & $0.076 \pm 0.043$ & $0.0004 \pm 0.0001$ & $0.0018$ \\
\cmidrule(lr){1-8}
\multicolumn{8}{@{}l}{\footnotesize\textit{NIC-only Spearman} ($n{=}8$):\quad$\rho_s(\tilde{\mathcal{L}},\cdot)=-0.98,-0.86,0.79$ for PSNR, MS-SSIM, and LPIPS, respectively ($p{=}3{\times}10^{-5},0.007,0.021$).} \\
\bottomrule
\end{tabular}
}
  \vspace{-1em}
  \end{table*}

\subsection{Leakage-Regularized Decoder Fine-Tuning}
\label{sec:example2}
\label{new:sec:finetune}
\label{new:sub:ft_joint}
\label{new:sub:ft_ablation}

\smallskip\noindent\textbf{Single-codec demonstration.}
We fine-tune only the Cheng2020-Anchor decoder on one $128{\times}128$
DCT-basis input, using the objective of Eq.~\eqref{eq:joint_objective}
with $\mathcal L_{\mathrm{dist}}{=}0$. After $900$ steps, median leakage
decreases from $0.291$ to $0.090$ ($69\%$) while low-frequency energy and
the bitstream remain unchanged (Fig.~\ref{fig:freq_analysis}(c)). Under the
standardized $1000$-step protocol used below, the same configuration
reaches $L^{\mathrm{med}}\approx0.001$
(Suppl.\ Table~\ref{new:tab:ft_basis}). Leakage can thus be minimized on
the training basis; a natural-image benefit does not follow.

\smallskip\noindent\textbf{Basis-only control.}
Extending this configuration (Eq.~\eqref{eq:joint_objective},
$\mathcal L_{\mathrm{dist}}{=}0$) to all eight architectures for $1000$
steps, held-out Kodak PSNR decreases by $10.6$-$22.0$\,dB for
the six older NICs and by $1.7$-$1.9$\,dB for TCM and FTIC
(Suppl.\ Table~\ref{new:tab:ft_basis}). The synthetic objective alone is
insufficient; the protocol below adds natural-image MSE.

\begin{table}[t]
  \centering
  \caption{Leakage-regularized decoder fine-tuning: DCT-basis leakage and
  Kodak quality, with an MSE-only control. $L^{\mathrm{med}}_{\mathrm{before/after}}$
  is the median leakage on the $256{\times}256$ composite basis; $\Delta$ metrics
  are reported on Kodak ($\Delta$: fine-tuned minus pretrained) with the
  bitstream fixed. Accepted
  operating points minimize $L^{\mathrm{med}}$ subject to $0.1$\,dB
  validation-PSNR limit. The MSE-only value is selected independently by the
  same rule. The BMSHJ2018-Factorized step-2000 endpoint exceeds this limit
  and is shown for completeness. Full protocol: Suppl.\
  Sec.~\ref{supp:ft_protocols}.}
  \label{new:tab:ft_joint}
  \smallskip
  \setlength{\tabcolsep}{3pt}
  \scriptsize
  \resizebox{\columnwidth}{!}{\begin{tabular}{lccccccc}
\toprule
& \multicolumn{6}{c}{MSE + leakage} & MSE-only control \\
\cmidrule(lr){2-7}\cmidrule(l){8-8}
Codec & $\lambda_{\mathrm{freq}}$ & step & $L^{\mathrm{med}}_{\mathrm{before}}$ & $L^{\mathrm{med}}_{\mathrm{after}}$ & $\Delta$PSNR & $\Delta$LPIPS & $L^{\mathrm{med}}_{\mathrm{after}}$ ($\Delta$PSNR) \\
\midrule
BMSHJ2018-Fact. & 0.003 & 2000 & 0.4558 & 0.0437 & -0.59 & $-5\mathrm{e}{-5}$ & 0.3812 (-0.02) \\
BMSHJ2018-Hyp. & 0.01 & 2000 & 0.1474 & 0.0034 & -0.26 & $+7\mathrm{e}{-4}$ & 0.1322 (-0.15) \\
MBT2018-Mean & 0.01 & 1500 & 0.1244 & 0.0042 & -0.29 & $+2.3\mathrm{e}{-3}$ & 0.1130 (-0.17) \\
MBT2018 & 0.003 & 2000 & 0.1151 & 0.0045 & -0.21 & $+2.4\mathrm{e}{-3}$ & 0.0867 (-0.11) \\
Cheng2020-Anchor & 0.03 & 500 & 0.7697 & 0.0012 & -0.17 & $+6\mathrm{e}{-4}$ & 0.1431 (-0.02) \\
Cheng2020-Attn. & 0.03 & 1750 & 0.3023 & 0.0006 & -0.26 & $+2.9\mathrm{e}{-3}$ & 0.1417 (-0.14) \\
TCM & 0.03 & 2000 & 0.0019 & 0.0001 & -0.09 & $-4\mathrm{e}{-4}$ & 0.0019 (+0.00) \\
FTIC & 0.03 & 1000 & 0.0044 & 0.0001 & -0.13 & $-1\mathrm{e}{-4}$ & 0.0044 (+0.00) \\
\bottomrule
\end{tabular}
}
\end{table}

\begin{figure*}[t]
  \centering
  \includegraphics[width=0.9\textwidth]{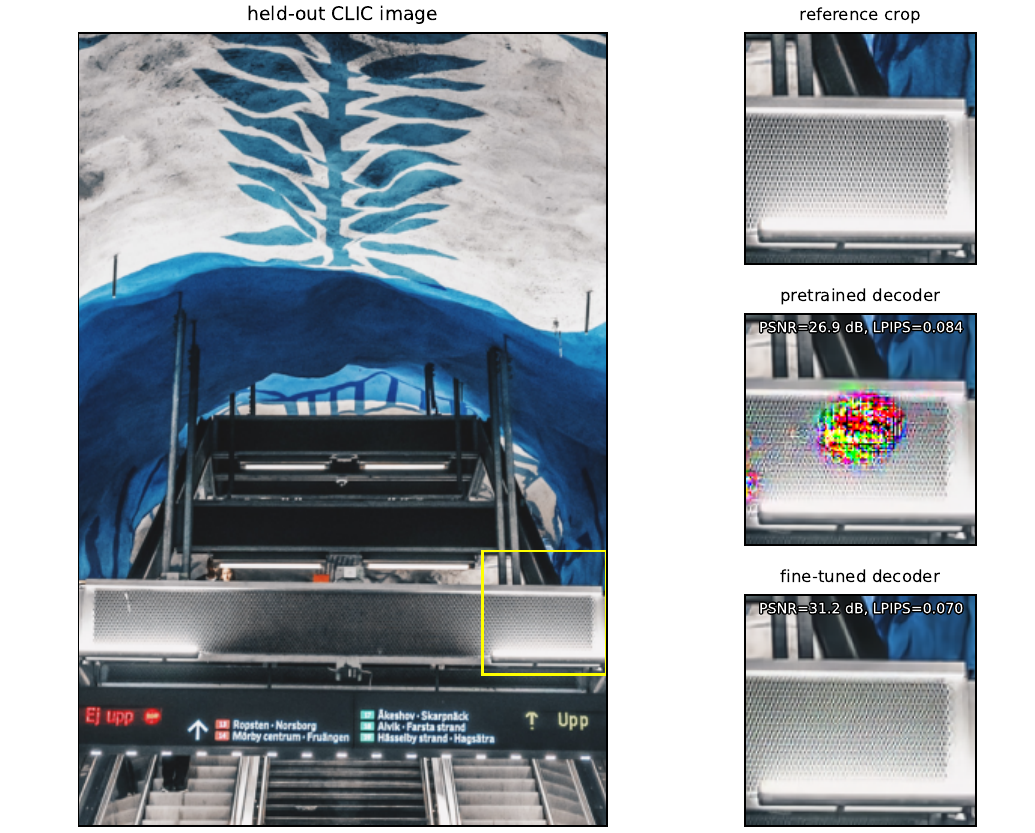}
  \caption{Leakage-regularized decoder fine-tuning on a held-out CLIC
  image. For Cheng2020-Anchor ($q{=}6$), the selected decoder removes the
  chromatic artifact from the same bitstream and raises PSNR from $26.9$ to
  $31.2$\,dB.}
  \label{new:fig:ft_natural}
\end{figure*}

\smallskip\noindent\textbf{Fine-tuning protocol.}
Table~\ref{new:tab:ft_joint} evaluates leakage-regularized decoder fine-tuning
across eight NIC architectures. Each fine-tuning step combines pixel MSE on a
batch of $4$ random crops from DIV2K ($256{\times}256$) with weighted leakage on the fixed
DCT basis matrix ($256{\times}256$) and two terms that preserve the
pretrained low-frequency response, as defined in
Eq.~\eqref{eq:joint_objective}. Fine-tuned checkpoints are evaluated at 250-step intervals up to
step 2,000. Among
the evaluated $(\lambda_{\mathrm{freq}},\mathrm{step})$ pairs, we select the point with the
lowest median basis leakage subject to an approximately $0.1$\,dB PSNR-loss
limit on eight held-out DIV2K images. After selection, the chosen decoder is evaluated on Kodak-24 at each
codec's native $q$ (rate reported from the entropy model);
Table~\ref{new:tab:ft_joint} reports mean $\Delta$PSNR and $\Delta$LPIPS
($\Delta$: fine-tuned minus pretrained). Sampling, codec settings, and the
evaluated $\lambda_{\mathrm{freq}}$ values are specified in
Suppl.\ Sec.~\ref{supp:ft_protocols}.

\smallskip\noindent\textbf{Results and control.}
For BMSHJ2018-Hyperprior, both MBT models, and both Cheng2020 models,
leakage-regularized fine-tuning reduces basis leakage to $0.0006$-$0.0045$,
compared with $0.087$-$0.143$ for independently selected MSE-only controls;
TCM and FTIC both reach approximately $10^{-4}$.
BMSHJ2018-Factorized also reduces leakage
($0.456\!\to\!0.044$), but its validation PSNR decreases by $0.416$\,dB,
beyond the $0.1$\,dB limit. Across the seven models accepted by the validation gate,
leakage-regularized fine-tuning reduces median DCT-basis leakage by factors of
$19$-$236$ relative to the MSE-only control. On Kodak, their PSNR decreases by
at most $0.29$\,dB; BMSHJ2018-Factorized decreases by $0.59$\,dB. For the
selected Cheng2020 checkpoints, median leakage also decreases on a separate
$512{\times}512$ basis: $0.73\!\to\!0.03$ for Anchor and
$0.30\!\to\!0.02$ for Attention. Full trajectories are reported in
Suppl.\ Fig.~\ref{new:fig:pareto}.

Despite this small average cost, decoder
fine-tuning removes localized chromatic artifacts from the same bitstream in
selected held-out examples (Fig.~\ref{new:fig:ft_natural}; Suppl.\
Figs.~\ref{sfig:gallery1}-\ref{sfig:gallery2}).

\section{Conclusion}

Controlled DCT inputs reveal frequency-dependent codec behavior that aggregate
rate-distortion scores do not capture. The basis-matrix probe measures
target-frequency gain and off-target redistribution under a joint stress
input, while isolated single-frequency probes provide an exact decomposition
of target-amplitude error and off-target energy with two-sided error bounds.

Across the evaluated codecs, the older BMSHJ, MBT, and Cheng2020 models show
substantially higher high-band off-target energy than the classical codecs,
whereas TCM and FTIC fall in the classical-codec range under this diagnostic.
The isolated probes also expose a frequency-specific pre-clamp instability in
released Cheng2020-Anchor checkpoints that scalar metrics do not reveal.

Relative off-target energy must be interpreted together with target-frequency
gain. In the evaluated transfer tests, synthetic response models did not
predict natural-image distortion quantitatively, and basis-only fine-tuning
did not preserve natural-image quality. The benchmark should therefore be used
as a controlled spectral diagnostic alongside standard natural-image
evaluation.

The same metric can also be used during decoder fine-tuning to penalize
reconstructed energy outside the target frequency. Among the seven accepted
models, this penalty reduced
off-target energy on the DCT probes by $94.7$-$99.8\%$ without changing
the encoded bitstream. Natural-image PSNR decreased by $0.09$-$0.29$\,dB,
while selected held-out examples showed removal of localized chromatic
artifacts. Extending the analysis to chroma and
temporal frequencies requires separate experiments.

\section*{CRediT authorship contribution statement}
\textbf{Nikolay I. Kalmykov:} Conceptualization, Methodology, Software,
Investigation, Visualization, Writing -- original draft.
\textbf{Razan Dibo:} Formal analysis.
\textbf{Yujun Du:} Formal analysis, Writing -- review \& editing.
\textbf{Yipeng Liu:} Supervision, Funding acquisition, Writing --
review \& editing.
\textbf{Ivan Oseledets:} Supervision, Funding acquisition.
\textbf{Anh-Huy Phan:} Conceptualization, Methodology, Validation, Supervision,
Project administration, Funding acquisition, Writing -- review \& editing.

\section*{Declaration of generative AI and AI-assisted technologies in the
manuscript preparation process}
During the preparation of this work the authors used ChatGPT (OpenAI) in
order to polish grammar and phrasing, suggest literature-search keywords,
and help debug minor Python issues in scripts. After using this service,
the authors reviewed and edited the content as needed and take full
responsibility for the content of the published article.

\bibliographystyle{elsarticle-num}
\bibliography{refs}

\begin{thebibliography}{10}
\expandafter\ifx\csname url\endcsname\relax
  \def\url#1{\texttt{#1}}\fi
\expandafter\ifx\csname urlprefix\endcsname\relax\def\urlprefix{URL }\fi
\expandafter\ifx\csname href\endcsname\relax
  \def\href#1#2{#2} \def\path#1{#1}\fi

\bibitem{kodakdataset}
Kodak, \href{https://r0k.us/graphics/kodak/}{Kodak lossless true color image
  suite} (1999).
\newline\urlprefix\url{https://r0k.us/graphics/kodak/}

\bibitem{clicdataset}
CLIC, \href{https://compression.cc}{Workshop and challenge on learned image
  compression ({CLIC})} (2020).
\newline\urlprefix\url{https://compression.cc}

\bibitem{div2kdataset}
E.~Agustsson, R.~Timofte, {NTIRE} 2017 challenge on single image
  super-resolution: Dataset and study, in: IEEE/CVF Conf. Computer Vision and
  Pattern Recognition Workshops (CVPRW), 2017, pp. 1122--1131.
\newblock \href {https://doi.org/10.1109/CVPRW.2017.150}
  {\path{doi:10.1109/CVPRW.2017.150}}.

\bibitem{Yang2022An}
Y.~Yang, S.~Mandt, L.~Theis, An introduction to neural data compression, Found.
  Trends Comput. Graph. Vis. 15 (2023) 113--200.
\newblock \href {https://doi.org/10.1561/0600000107}
  {\path{doi:10.1561/0600000107}}.

\bibitem{ma2019image}
S.~Ma, X.~Zhang, C.~Jia, Z.~Zhao, S.~Wang, S.~Wang, Image and video compression
  with neural networks: A review, IEEE Transactions on Circuits and Systems for
  Video Technology 30 (2020) 1683--1698.
\newblock \href {https://doi.org/10.1109/tcsvt.2019.2910119}
  {\path{doi:10.1109/tcsvt.2019.2910119}}.

\bibitem{lieberman2023spectral}
K.~Lieberman, J.~Diffenderfer, C.~Godfrey, B.~Kailkhura, Neural image
  compression: Generalization, robustness, and spectral biases, in: Advances in
  Neural Information Processing Systems (NeurIPS), Vol.~36, 2023, pp.
  77201--77241.

\bibitem{balle2017endtoend}
J.~Ball{\'e}, V.~Laparra, E.~P. Simoncelli,
  \href{https://openreview.net/forum?id=rJxdQ3jeg}{End-to-end optimized image
  compression}, in: International Conference on Learning Representations
  (ICLR), 2017.
\newline\urlprefix\url{https://openreview.net/forum?id=rJxdQ3jeg}

\bibitem{cheng2020attention}
Z.~Cheng, H.~Sun, M.~Takeuchi, J.~Katto, Learned image compression with
  discretized gaussian mixture likelihoods and attention mechanism, in:
  IEEE/CVF Conference on Computer Vision and Pattern Recognition (CVPR), 2020,
  pp. 7939--7948.

\bibitem{Liu_2023_CVPR}
J.~Liu, H.~Sun, J.~Katto, Learned image compression with mixed
  transformer-{CNN} architectures, in: IEEE/CVF Conf. Computer Vision and
  Pattern Recognition (CVPR), 2023, pp. 14388--14397.

\bibitem{mentzer2020hific}
F.~Mentzer, G.~Toderici, M.~Tschannen, E.~Agustsson,
  \href{https://papers.neurips.cc/paper_files/paper/2020/file/8a50bae297807da9e97722a0b3fd8f27-Paper.pdf}{High-fidelity
  generative image compression}, in: Advances in Neural Information Processing
  Systems (NeurIPS), 2020.
\newline\urlprefix\url{https://papers.neurips.cc/paper_files/paper/2020/file/8a50bae297807da9e97722a0b3fd8f27-Paper.pdf}

\bibitem{he2022elic}
D.~He, Z.~Yang, W.~Peng, R.~Ma, H.~Qin, Y.~Wang, {ELIC}: Efficient learned
  image compression with unevenly grouped space-channel contextual adaptive
  coding, in: IEEE/CVF Conf. Computer Vision and Pattern Recognition (CVPR),
  2022, pp. 5718--5727.

\bibitem{Chen2020Variable}
T.~Chen, Z.~Ma, Variable bitrate image compression with quality scaling
  factors, in: Proc. IEEE Int. Conf. Acoustics, Speech, and Signal Processing
  (ICASSP), 2020, pp. 2163--2167.
\newblock \href {https://doi.org/10.1109/icassp40776.2020.9053885}
  {\path{doi:10.1109/icassp40776.2020.9053885}}.

\bibitem{wang2004ssim}
Z.~Wang, A.~C. Bovik, H.~R. Sheikh, E.~P. Simoncelli, Image quality assessment:
  From error visibility to structural similarity, IEEE Transactions on Image
  Processing 13~(4) (2004) 600--612.

\bibitem{blau2019rdp}
Y.~Blau, T.~Michaeli, Rethinking lossy compression: The
  rate-distortion-perception tradeoff, in: Int. Conf. Machine Learning (ICML),
  2019, pp. 675--685.

\bibitem{huang2015urban100}
J.-B. Huang, A.~Singh, N.~Ahuja, Single image super-resolution from transformed
  self-exemplars, in: IEEE/CVF Conf. Computer Vision and Pattern Recognition
  (CVPR), 2015, pp. 5197--5206.

\bibitem{cimpoi2014dtd}
M.~Cimpoi, S.~Maji, I.~Kokkinos, S.~Mohamed, A.~Vedaldi, Describing textures in
  the wild, in: IEEE/CVF Conf. Computer Vision and Pattern Recognition (CVPR),
  2014, pp. 3606--3613.

\bibitem{ahmed1974dct}
N.~Ahmed, T.~Natarajan, K.~R. Rao, Discrete cosine transform, IEEE Transactions
  on Computers C-23~(1) (1974) 90--93.

\bibitem{wallace1991jpeg}
G.~K. Wallace, The {JPEG} still picture compression standard, Communications of
  the {ACM} 34~(4) (1991) 30--44.
\newblock \href {https://doi.org/10.1145/103085.103089}
  {\path{doi:10.1145/103085.103089}}.

\bibitem{sullivan2012hevc}
G.~J. Sullivan, J.-R. Ohm, W.-J. Han, T.~Wiegand, Overview of the high
  efficiency video coding ({HEVC}) standard, IEEE Transactions on Circuits and
  Systems for Video Technology 22~(12) (2012) 1649--1668.

\bibitem{sun2020dct}
M.~Sun, X.~He, S.~Xiong, C.~Ren, X.~Li, Reduction of {JPEG} compression
  artifacts based on {DCT} coefficients prediction, Neurocomputing 384 (2020)
  335--345.
\newblock \href {https://doi.org/10.1016/j.neucom.2019.12.015}
  {\path{doi:10.1016/j.neucom.2019.12.015}}.

\bibitem{balle2018hyperprior}
J.~Ball{\'e}, D.~Minnen, S.~Singh, S.~J. Hwang, N.~Johnston,
  \href{https://openreview.net/forum?id=rkcQFMZRb}{Variational image
  compression with a scale hyperprior}, in: International Conference on
  Learning Representations (ICLR), 2018.
\newline\urlprefix\url{https://openreview.net/forum?id=rkcQFMZRb}

\bibitem{minnen2018joint}
D.~Minnen, J.~Ball{\'e}, G.~D. Toderici,
  \href{https://papers.nips.cc/paper_files/paper/2018/file/53edebc543333dfbf7c5933af792c9c4-Paper.pdf}{Joint
  autoregressive and hierarchical priors for learned image compression}, in:
  Advances in Neural Information Processing Systems (NeurIPS), 2018.
\newline\urlprefix\url{https://papers.nips.cc/paper_files/paper/2018/file/53edebc543333dfbf7c5933af792c9c4-Paper.pdf}

\bibitem{mishra2022attention}
D.~Mishra, S.~K. Singh, R.~K. Singh, Deep {CNN} based image compression with
  redundancy minimization via attention guidance, Neurocomputing 507 (2022)
  397--411.
\newblock \href {https://doi.org/10.1016/j.neucom.2022.08.009}
  {\path{doi:10.1016/j.neucom.2022.08.009}}.

\bibitem{alshina2024jpegai}
E.~Alshina, J.~Ascenso, T.~Ebrahimi, {JPEG AI}: The first international
  standard for image coding based on an end-to-end learning-based approach,
  IEEE MultiMedia 31~(4) (2024) 60--69.
\newblock \href {https://doi.org/10.1109/MMUL.2024.3485255}
  {\path{doi:10.1109/MMUL.2024.3485255}}.

\bibitem{Hu2020Learning}
Y.~Hu, W.~Yang, Z.~Ma, J.~Liu, Learning end-to-end lossy image compression: A
  benchmark, IEEE Transactions on Pattern Analysis and Machine Intelligence 44
  (2022) 4194--4211.
\newblock \href {https://doi.org/10.1109/tpami.2021.3065339}
  {\path{doi:10.1109/tpami.2021.3065339}}.

\bibitem{rahaman2019spectral}
N.~Rahaman, A.~Baratin, D.~Arpit, F.~Draxler, M.~Lin, F.~A. Hamprecht,
  Y.~Bengio, A.~Courville,
  \href{https://proceedings.mlr.press/v97/rahaman19a.html}{On the spectral bias
  of neural networks}, in: Proceedings of the 36th International Conference on
  Machine Learning, Vol.~97 of Proceedings of Machine Learning Research, PMLR,
  2019, pp. 5301--5310.
\newline\urlprefix\url{https://proceedings.mlr.press/v97/rahaman19a.html}

\bibitem{gao2021frequency}
G.~Gao, P.~You, R.~Pan, S.~Han, Y.~Zhang, Y.~Dai, H.~Lee,
  \href{https://openaccess.thecvf.com/content/ICCV2021/html/Gao_Neural_Image_Compression_via_Attentional_Multi-Scale_Back_Projection_and_Frequency_ICCV_2021_paper.html}{Neural
  image compression via attentional multi-scale back projection and frequency
  decomposition}, in: Proceedings of the IEEE/CVF International Conference on
  Computer Vision (ICCV), 2021, pp. 14677--14686.
\newline\urlprefix\url{https://openaccess.thecvf.com/content/ICCV2021/html/Gao_Neural_Image_Compression_via_Attentional_Multi-Scale_Back_Projection_and_Frequency_ICCV_2021_paper.html}

\bibitem{li2024ftic}
H.~Li, S.~Li, W.~Dai, C.~Li, J.~Zou, H.~Xiong,
  \href{https://openreview.net/forum?id=HKGQDDTuvZ}{Frequency-aware transformer
  for learned image compression}, in: International Conference on Learning
  Representations (ICLR), 2024.
\newline\urlprefix\url{https://openreview.net/forum?id=HKGQDDTuvZ}

\bibitem{fu2024weconvene}
H.~Fu, J.~Liang, Z.~Fang, J.~Han, F.~Liang, G.~Zhang, {WeConvene}: Learned
  image compression with wavelet-domain convolution and entropy model, in:
  Computer Vision -- ECCV 2024, Vol. 15108 of Lecture Notes in Computer
  Science, Springer, 2024, pp. 37--53.
\newblock \href {https://doi.org/10.1007/978-3-031-72973-7_3}
  {\path{doi:10.1007/978-3-031-72973-7_3}}.

\bibitem{li2025disentangled}
H.~Li, S.~Li, W.~Dai, M.~Cao, N.~Kan, C.~Li, J.~Zou, H.~Xiong, On disentangled
  training for nonlinear transform in learned image compression, in:
  International Conference on Learning Representations (ICLR), 2025.

\bibitem{ma2026adaptivefrequency}
H.~Ma, X.~Shi, H.~Sun, X.~Yue, X.~Liu, G.~Wang, W.~Cai,
  \href{https://openaccess.thecvf.com/content/CVPR2026/html/Ma_Learned_Image_Compression_via_Sparse_Attention_and_Adaptive_Frequency_CVPR_2026_paper.html}{Learned
  image compression via sparse attention and adaptive frequency}, in:
  Proceedings of the IEEE/CVF Conference on Computer Vision and Pattern
  Recognition (CVPR), 2026, pp. 41278--41287.
\newline\urlprefix\url{https://openaccess.thecvf.com/content/CVPR2026/html/Ma_Learned_Image_Compression_via_Sparse_Attention_and_Adaptive_Frequency_CVPR_2026_paper.html}

\bibitem{cong2026spectralregularization}
W.~Cong, J.~Shi, M.~Lu, X.~Zhang, Z.~Ma,
  \href{https://iclr.cc/virtual/2026/poster/10007684}{Taming hierarchical image
  coding optimization: A spectral regularization perspective}, in:
  International Conference on Learning Representations (ICLR), 2026.
\newline\urlprefix\url{https://iclr.cc/virtual/2026/poster/10007684}

\bibitem{Chen2021Toward}
T.~Chen, Z.~Ma, Toward robust neural image compression: Adversarial attack and
  model finetuning, IEEE Transactions on Circuits and Systems for Video
  Technology 33 (2023) 7842--7856.
\newblock \href {https://doi.org/10.1109/tcsvt.2023.3276442}
  {\path{doi:10.1109/tcsvt.2023.3276442}}.

\bibitem{liu2024malice}
K.~Liu, D.~Wu, Y.~Wu, Y.~Wang, D.~Feng, B.~Tan, S.~Garg, Manipulation attacks
  on learned image compression, IEEE Transactions on Artificial Intelligence
  5~(6) (2024) 3083--3097.
\newblock \href {https://doi.org/10.1109/TAI.2023.3340982}
  {\path{doi:10.1109/TAI.2023.3340982}}.

\bibitem{kalmykov2026t}
N.~I. Kalmykov, R.~Dibo, K.~Shen, Z.~Xu, A.-H. Phan, Y.~Liu, I.~Oseledets,
  {T-MLA}: A targeted multiscale log--exponential attack framework for neural
  image compression, Information Sciences 738 (2026) 123143.
\newblock \href {https://doi.org/10.1016/j.ins.2026.123143}
  {\path{doi:10.1016/j.ins.2026.123143}}.

\bibitem{bychkov2025nicrobustbench}
G.~Bychkov, K.~Abud, E.~Kovalev, A.~Gushchin, S.~Lavrushkin, D.~Vatolin,
  A.~Antsiferova, {NIC-RobustBench}: A comprehensive open-source toolkit for
  neural image compression and robustness analysis, arXiv preprint
  arXiv:2506.19051 (2025).

\bibitem{kim2022successive}
J.-H. Kim, S.~Jang, J.-H. Choi, J.-S. Lee, Successive learned image
  compression: Comprehensive analysis of instability, Neurocomputing 506 (2022)
  12--24.
\newblock \href {https://doi.org/10.1016/j.neucom.2022.07.065}
  {\path{doi:10.1016/j.neucom.2022.07.065}}.

\bibitem{muckley2023illm}
M.~J. Muckley, A.~El-Nouby, K.~Ullrich, H.~J{\'e}gou, J.~Verbeek,
  \href{https://proceedings.mlr.press/v202/muckley23a.html}{Improving
  statistical fidelity for neural image compression with implicit local
  likelihood models}, in: Proceedings of the 40th International Conference on
  Machine Learning, Vol. 202 of Proceedings of Machine Learning Research, PMLR,
  2023, pp. 25426--25443.
\newline\urlprefix\url{https://proceedings.mlr.press/v202/muckley23a.html}

\bibitem{yang2023conditionaldiffusion}
R.~Yang, S.~Mandt, Lossy image compression with conditional diffusion models,
  in: Advances in Neural Information Processing Systems, Vol.~36, 2023, pp.
  64971--64995.

\bibitem{jiang2026sscl}
W.~Jiang, Y.~Zhai, J.~Yang, B.~Feng, W.~Wang, B.~Huang, L.~Ding, R.~Wang,
  {SSCL}: Adversarially guided image compression via semantic and spectral
  consistency learning, in: Proceedings of the AAAI Conference on Artificial
  Intelligence, Vol.~40, 2026, pp. 14937--14945.
\newblock \href {https://doi.org/10.1609/aaai.v40i17.38515}
  {\path{doi:10.1609/aaai.v40i17.38515}}.

\bibitem{begaint2020compressai}
J.~B{\'e}gaint, F.~Racap{\'e}, S.~Feltman, A.~Pushparaja, {CompressAI}: A
  {PyTorch} library and evaluation platform for end-to-end compression
  research, arXiv preprint arXiv:2011.03029 (2020).

\bibitem{mannos1974}
J.~L. Mannos, D.~J. Sakrison, The effects of a visual fidelity criterion on the
  encoding of images, IEEE Transactions on Information Theory 20~(4) (1974)
  525--536.
\newblock \href {https://doi.org/10.1109/TIT.1974.1055250}
  {\path{doi:10.1109/TIT.1974.1055250}}.

\end{thebibliography}

\setcounter{figure}{0}
\setcounter{table}{0}
\setcounter{equation}{0}
\setcounter{theorem}{0}
\setcounter{section}{0}
\setcounter{subsection}{0}
\renewcommand{\thesection}{\Alph{section}}
\renewcommand{\thesubsection}{\thesection.\arabic{subsection}}
\renewcommand{\thefigure}{S\arabic{figure}}
\renewcommand{\thetable}{S\arabic{table}}
\renewcommand{\theequation}{S\arabic{equation}}
\renewcommand{\theHequation}{S\arabic{equation}}
\renewcommand{\thetheorem}{S\arabic{theorem}}
\renewcommand{\theHtheorem}{S\arabic{theorem}}
\renewcommand{\theHsection}{supp.\Alph{section}}
\renewcommand{\theHsubsection}{supp.\Alph{section}.\arabic{subsection}}
\renewcommand{\theHfigure}{supp.S\arabic{figure}}
\renewcommand{\theHtable}{supp.S\arabic{table}}

\twocolumn[
{\centering
\large\bfseries Supplementary Material\par
\vspace{0.25em}
\normalsize\normalfont Benchmarking Neural Image Compression with DCT-Based Spectral Diagnostics\par
\vspace{0.65em}}
]
\renewcommand{\topfraction}{0.92}
\renewcommand{\bottomfraction}{0.75}
\renewcommand{\textfraction}{0.06}
\renewcommand{\dbltopfraction}{0.92}
\renewcommand{\floatpagefraction}{0.80}
\renewcommand{\dblfloatpagefraction}{0.80}
\setcounter{topnumber}{3}
\setcounter{bottomnumber}{2}
\setcounter{totalnumber}{5}
\setcounter{dbltopnumber}{3}

\begin{table}[t!]
\centering
\caption{Core notation used in the main paper and Supplementary Material.}
\label{tab:supp_notation}
\setlength{\tabcolsep}{3pt}
\renewcommand{\arraystretch}{1.02}
\footnotesize
\begin{tabular}{@{}p{0.26\columnwidth}p{0.715\columnwidth}@{}}
\toprule
\textbf{Notation} & \textbf{Definition} \\
\midrule
$n,\ k$ & DCT size and frequency index, $k\in\{0,\ldots,n-1\}$. \\
$\mathbf D,\ \mathbf d_k$ & Orthonormal DCT-basis image and its $k$-th column. \\
$\mathcal D,\ \mathcal D_2$ & Orthonormal one- and two-dimensional DCT operators. \\
$\widehat{\mathbf d}_k,\ \widehat{\mathbf X}$ & Reconstructed basis-matrix column and isolated probe. \\
$R_{i,k}$ & Fraction of reconstructed column $k$'s DCT energy located in bin $i$. \\
$L_k,\ g_k$ & Relative off-target energy and signed target-frequency gain. \\
$\mathbf X_k^{(1)},\ \mathbf X_k^{(2)}$ & Unit-energy isolated probes ($\mathbf X$ denotes either); target bins $(k,0)$, $(k,k)$. \\
$s$ & Amplitude scale of the centered embedding $x\mapsto\tfrac12+sx$. \\
$E_k^{\mathrm{off}},\ \tau_k,\ \lambda_k$ & Off-bin energy, total reconstructed spectral energy, and isolated-probe relative leakage. \\
$\mu_k,\Delta_k,\sigma_k^{\mathrm{norm}},H_k^{\mathrm{norm}}$ & Spectral centroid, normalized shift, spread, and entropy. \\
$\mathrm{ODR}^{\mathrm{norm}}_k$ & Bounded off-/on-target energy ratio, monotone in $L_k$. \\
$\bar L,\ L^{\mathrm{med}},\ L_{\mathrm{low}/\mathrm{high}}$ & Mean, median, and low/high-band means of $L_k$. \\
$\rho(f),\ \bar\rho$ & Natural-image radial distortion fraction and its mean over frequency bands. \\
$\widetilde{\mathcal L},\ \widetilde{\mathcal L}/\bar\rho$ & Descriptive synthetic-natural spectral alignment and its magnitude-normalized form. \\
$\mathbf c_k,\ \delta_k$ & Row $k$ of the column-DCT coefficients and its exact squared-error share. \\
$T_{ik},\ \alpha_k$ & Measured linear response operator and basis-measured attenuation. \\
$\mathcal L_{\mathrm{freq}},\ \lambda_{\mathrm{freq}}$ & Leakage regularizer and its weight in decoder fine-tuning. \\
$w_k,\ k_0,\ p$ & Weights of $\mathcal L_{\mathrm{freq}}$, their cutoff index, and exponent. \\
$\mathcal L_{\mathrm{dist}},\ \mathcal G_{\mathrm{pur}},\ \mathcal G_{\mathrm{mse}}$ & Natural-image MSE term; low-band purity/MSE guards. \\
$q,\ \mathrm{bpp}$ & Checkpoint/quality index and rate in bits per pixel. \\
$\epsilon$ & Numerical constant ($10^{-12}$). \\
\bottomrule
\end{tabular}
\end{table}

\section{Notation and Definitions}
Table~\ref{tab:supp_notation} collects the notation used throughout the
main paper and Supplementary Material.

\section{Probe Construction and Mathematical Details}

\subsection{DCT Preliminaries}
\label{sec:supp_dct_prelim}

\smallskip\noindent\textbf{Basis.}
The orthonormal DCT-II basis matrix $\mathbf D\in\mathbb R^{n\times n}$ has
entries
\begin{equation}
\mathbf D[m,k]=\alpha_k\cos\Bigl[\tfrac{\pi}{n}\bigl(m+\tfrac12\bigr)k\Bigr],
\qquad
\alpha_0=\sqrt{\tfrac1n},\quad \alpha_{k>0}=\sqrt{\tfrac2n},
\label{eq:supp_dct_basis}
\end{equation}
and satisfies $\mathbf D^\top\mathbf D=\mathbf I$. Its $k$-th column
$\mathbf d_k$ is the $k$-th basis wave and has unit norm.

\smallskip\noindent\textbf{Basis, transform, and probe.}
The bold $\mathbf D$ is the matrix of Eq.~\eqref{eq:supp_dct_basis}, used both
as a change of basis and, read as an image, as the basis-matrix probe. The script $\mathcal D(\mathbf v)=\mathbf D^\top\mathbf v$ is the
one-dimensional transform of a single column, used to build the response
matrix $R$. The script $\mathcal D_2(\mathbf X)=\mathbf D^\top\mathbf X\mathbf D$
is the two-dimensional transform, i.e.\ $\mathcal D$ applied to every column
and then to every row. Orthonormality gives
\begin{equation}
\mathcal D_2(\mathbf D)=\mathbf D^\top\mathbf D\,\mathbf D=\mathbf D ,
\end{equation}
so the basis-matrix probe is a fixed point of the 2-D DCT: analyzing it in two
dimensions returns the probe itself and carries no per-frequency information.
The basis-matrix probe is therefore analyzed columnwise with $\mathcal D$,
while the isolated probes of Sec.~\ref{sec:supp_singlefreq}, which are single
2-D components by construction, are analyzed with $\mathcal D_2$.

\smallskip\noindent\textbf{Energy.}
Both transforms are orthonormal, so
$\|\mathcal D(\mathbf v)\|_2=\|\mathbf v\|_2$ and
$\|\mathcal D_2(\mathbf X)\|_F=\|\mathbf X\|_F$ (Parseval). The error
decompositions and bounds of Sec.~\ref{sec:supp_metric_bounds} rest on this
equality of spatial and spectral energy.

\smallskip\noindent\textbf{Index and frequency.}
Index $k$ corresponds to the spatial frequency $f_k=k/(2n)$ in cycles per
pixel: $k{=}0$ is the DC term and the Nyquist limit $f{=}0.5$ is reached at
$k{=}n$, so the sampled range $k\le n-1$ spans the full band. This mapping is used by the
perceptual-weighting check of Sec.~\ref{new:sub:csf_weight}.

\subsection{Diagonal of the DCT Basis as a Chirp Signal}
\label{sec:supp_chirp}

Define the diagonal $s_k = \mathbf{D}[k,k]$, $k=0,\dots,n-1$. Then $s_k$ is a discrete chirp-like signal whose effective instantaneous frequency $f_k^{\mathrm{eff}}$ rises and falls periodically, with turning points near $k \approx n/2$, caused by Nyquist folding of its quadratic phase.

\begin{proof}
Setting $m=k$ in Eq.~\eqref{eq:supp_dct_basis} gives
$s_k=\alpha_k\cos\Bigl[\tfrac{\pi}{n}\bigl(k^2+\tfrac{k}{2}\bigr)\Bigr]$
with phase $\phi_k=\tfrac{\pi}{n}\bigl(k^2+\tfrac{k}{2}\bigr)$.
The forward difference yields the instantaneous angular frequency
\[
\omega_k=\phi_{k+1}-\phi_k=\frac{\pi}{n}\!\left(2k+\tfrac{3}{2}\right),
\quad
f_k=\frac{\omega_k}{2\pi}=\frac{2k+1.5}{2n}.
\]
Because $\phi_k$ is quadratic, $f_k$ grows linearly with $k$, making $s_k$ chirp-like. Whenever $f_k$ exceeds the Nyquist limit $0.5$\,cycles/sample, the real cosine aliases it; writing $\tilde f_k=f_k\bmod 1$ for the aliased frequency,
\[
f_k^{\mathrm{eff}}=\min\bigl(\tilde f_k,\;1-\tilde f_k\bigr)\in\bigl[0,\tfrac12\bigr].
\]
Evaluating the analytic phase extension at key indices gives
$f_0=\frac{3}{4n}$, $f_{n/2}=\tfrac12+f_0$, and $f_n=1+f_0$; hence the
effective frequency approaches its trough at the end of the sampled diagonal,
while near $k\approx n/2$ it reaches the Nyquist-folded peak
$\tfrac12-f_0$.
Between folding points $f_k^{\mathrm{eff}}$ rises linearly; after each fold it falls, producing the periodic rise-fall pattern along the diagonal.
\end{proof}

\subsection{Single-Frequency Construction Details}
\label{sec:supp_singlefreq}

\begin{figure}[t]
  \centering
  \includegraphics[width=0.94\columnwidth]{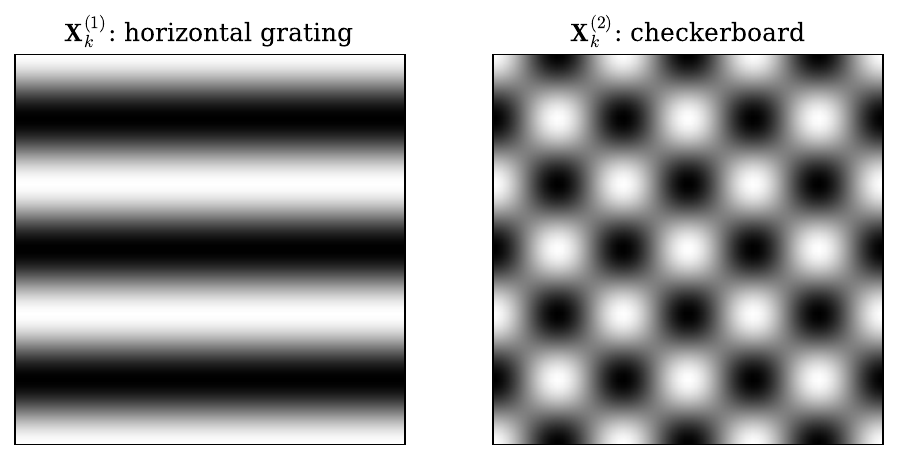}
  \caption{Isolated probes ($n{=}64$, $k{=}6$). $\mathbf X^{(1)}_k$ repeats
  one basis wave along every column: a horizontal grating with all spectral
  energy in bin $(k,0)$. $\mathbf X^{(2)}_k$ is separable in both axes: a
  checkerboard with all energy in bin $(k,k)$.}
  \label{sf:fig:probe_construction}
\end{figure}

\smallskip\noindent\textbf{Construction.}
Both probes of Eqs.~\eqref{sf:eq:X1}-\eqref{sf:eq:X2} are outer products, so
their entries are read off directly:
\[
  \mathbf X_k^{(1)}[m,j]=n^{-1/2}\,\mathbf d_k[m],
  \qquad
  \mathbf X_k^{(2)}[m,j]=\mathbf d_k[m]\,\mathbf d_k[j].
\]
Every column of $\mathbf X_k^{(1)}$ is the same wave $\mathbf d_k$, so the
image is constant along rows and varies only down the columns: a horizontal
grating. $\mathbf X_k^{(2)}$ varies along both axes and forms a checkerboard
(Fig.~\ref{sf:fig:probe_construction}). The factor $n^{-1/2}$ normalizes the
first probe, since $\|\mathbf d_k\mathbf 1^\top\|_F=\|\mathbf d_k\|_2\|\mathbf
1\|_2=\sqrt n$, whereas $\mathbf d_k\mathbf d_k^\top$ already has unit norm.

\smallskip\noindent\textbf{Amplitude and error scales.}
The embedded stimulus $\tfrac12+s\mathbf X$ is quoted either through its peak
amplitude $a=s\max_{m,j}|\mathbf X[m,j]|$, where $a{=}0.45$ places the
stimulus in $[0.05,0.95]$ around mid-grey, or through the raw scale $s$
applied to the unit-energy probe, whose root-mean-square entry is $1/n$. The
two conventions differ by an $n$-dependent factor and are stated per suite in
Sec.~\ref{sec:supp_repro}. Unit input energy also fixes the scale of the
squared error $e^2=\|\mathbf X-\widehat{\mathbf X}\|_F^2$ of
Eq.~\eqref{sf:eq:exact}: $e^2{=}0$ is exact reconstruction, $e^2{\approx}1$
corresponds to discarding the stimulus and returning a flat field,
and $e^2{>}1$ means the output carries more error energy than the input
contained. The divergence threshold used in Sec.~\ref{sec:supp_x1_gallery} is
$e^2{>}10$.

\smallskip\noindent\textbf{Centered embedding check.}
The isolated probes use $x\mapsto\tfrac12+sx$ with a fixed,
$k$-independent scale. In a controlled simulation, a pure $10\%$ attenuation
passed through a per-image min-max map produces a spurious, $k$-dependent
Frobenius error several times the true value, whereas the centered map
recovers the injected $0.10$ exactly. The centered map is therefore required to keep the stimulus amplitude
and DC level independent of frequency.

\smallskip\noindent\textbf{Single-bin property.}
For the constructions in Eqs.~\eqref{sf:eq:X1}-\eqref{sf:eq:X2},
$\mathbf D^\top\mathbf d_k=\mathbf e_k$ by orthonormality and
$\mathbf 1^\top\mathbf D=\sqrt n\,\mathbf e_0^\top$ because
$\mathbf d_0=n^{-1/2}\mathbf 1$. Hence
\begin{align}
\mathcal D_2(\mathbf X_k^{(1)})
&=n^{-1/2}(\mathbf D^\top\mathbf d_k)(\mathbf 1^\top\mathbf D)
=\mathbf e_k\mathbf e_0^\top,\\
\mathcal D_2(\mathbf X_k^{(2)})
&=(\mathbf D^\top\mathbf d_k)(\mathbf d_k^\top\mathbf D)
=\mathbf e_k\mathbf e_k^\top.
\end{align}
Thus each probe carries one unit of energy in exactly one 2-D DCT bin before
compression, which makes reconstruction errors comparable across probes and
across frequencies.

\subsection{Metric Identities and Leakage-Bound Proofs}
\label{sec:supp_metric_bounds}

All exact relationships below were checked on every evaluated codec run; the
largest numerical residual is below $10^{-6}$.

\smallskip\noindent\textbf{Attenuation vs.\ redistribution.}
Pure attenuation, $\widehat{\mathbf d}_k=\alpha\mathbf d_k$, gives
$L_k{=}0$ and $g_k{=}\alpha$: leakage alone would report a perfect response.
Pure redistribution at preserved target amplitude,
$\widehat{\mathbf d}_k=\mathbf d_k+\sum_{i\ne k}c_i\mathbf d_i$, gives
$g_k{=}1$ and $L_k=\sum_{i\ne k}c_i^2/(1+\sum_{i\ne k}c_i^2)>0$. Neither
quantity determines the other, which is why both are reported.

\smallskip\noindent\textbf{ODR and leakage.}
Each positive-energy column of $R$ sums to one, so
$\sum_{i\ne k}R_{i,k}=L_k$ and
\begin{equation}
\mathrm{ODR}^{\mathrm{norm}}_k
=\tanh\!\left[\frac{L_k}{2(1-L_k)+2\epsilon}\right].
\end{equation}
ODR is therefore a strictly monotone reparameterization of leakage (maximum
identity residual $9{\cdot}10^{-12}$ across the 11 codecs). In finite
precision, saturation of $\tanh$ creates ties above $L_k\gtrsim0.97$.

\smallskip\noindent\textbf{Pooled auxiliary correlations.}
Pooled over the eight learned codecs at their nearest-1\,bpp configurations
of Fig.~\ref{fig:ex1_dct_grid_allmodels} (2048 frequency-codec pairs),
$\rho_s(L_k,H_k^{\mathrm{norm}})=0.98$ and
$\rho_s(L_k,\sigma_k^{\mathrm{norm}})=0.92$. Centroid shift is the most
distinct auxiliary descriptor, although it remains correlated with leakage
($\rho_s(L_k,|\Delta_k|)=0.81$).

\smallskip\noindent\textbf{Parallel-axis identity.}
In raw index units, the second moment about the target decomposes as
\begin{equation}
\sum_i(i-k)^2R_{i,k}
=\underbrace{\sum_i(i-\mu_k)^2R_{i,k}}_{\sigma_k^2}
+(\mu_k-k)^2.
\end{equation}
Thus $\sigma_k$ measures symmetric dispersion and $\mu_k{-}k$ coherent
displacement. The normalized $\sigma_k^{\mathrm{norm}}$ and $\Delta_k$ of
Sec.~\ref{sec:dct_analysis} use different denominators, so the exact identity
applies to the raw index-domain quantities.

\smallskip\noindent\textbf{Single-frequency decomposition and bounds.}
Parseval's identity for the orthonormal 2-D DCT gives
\begin{equation}
\|\mathbf X-\widehat{\mathbf X}\|_F^2
=(1-g_k)^2+E_k^{\mathrm{off}},
\end{equation}
because the input spectrum has one unit entry and all other coefficients are
zero. Write
$\tau_k=\|\widehat{\mathbf G}\|_F^2=g_k^2+E_k^{\mathrm{off}}>0$. Then
\begin{align}
\|\mathbf X-\widehat{\mathbf X}\|_F^2-\lambda_k
&=\tau_k-2g_k+\frac{g_k^2}{\tau_k}\\
&=\frac{(\tau_k-g_k)^2}{\tau_k}
=\frac{(E_k^{\mathrm{off}}-g_k(1-g_k))^2}{\tau_k}\ge0.
\end{align}
This proves the lower bound in Eq.~\eqref{sf:eq:bounds_lower}; equality holds
iff $E_k^{\mathrm{off}}=g_k(1-g_k)$. If $\tau_k=1$ and $g_k\ge0$, then
$g_k=\sqrt{1-\lambda_k}$ and
\begin{equation}
\|\mathbf X-\widehat{\mathbf X}\|_F^2
=2\bigl(1-\sqrt{1-\lambda_k}\bigr)\le2\lambda_k,
\end{equation}
which proves Eq.~\eqref{sf:eq:bounds_upper}. The constants are attained, so
they are optimal.

\smallskip\noindent\textbf{Full-basis bounds.}
For column $k$, let
$E_k=\|\mathcal D(\widehat{\mathbf d}_k)\|^2>0$ and
$a_k=\mathcal D(\widehat{\mathbf d}_k)_k$, so
$a_k^2=R_{k,k}E_k$. Parseval gives
$\|\widehat{\mathbf d}_k-\mathbf d_k\|^2=E_k-2a_k+1$.
If $a_k<0$, this is at least $1\ge L_k$. If $a_k\ge0$, substitute
$a_k=\sqrt{R_{k,k}E_k}$; minimizing over $E_k>0$ gives
$E_k=R_{k,k}$ and minimum $1-R_{k,k}=L_k$. Averaging proves
Eq.~\eqref{new:eq:frob_lower}. Under $E_k=1$ and $a_k\ge0$,
\begin{equation}
\|\widehat{\mathbf d}_k-\mathbf d_k\|^2
=2\bigl(1-\sqrt{R_{k,k}}\bigr)\le2(1-R_{k,k})=2L_k,
\end{equation}
and averaging proves Eq.~\eqref{new:eq:frob_upper}. The energy-preservation
condition is substantive: the released Cheng2020-Anchor checkpoints violate
it on unstable isolated inputs, so the upper bound is not unconditional.

\FloatBarrier
\section{Secondary Diagnostic Checks}

\subsection{Perceptual-Weighting Robustness Check}
\label{new:sub:csf_weight}

To test the equal weighting of frequencies, let $f_k=k/(2n)$ cycles per pixel and
let $v$ be the viewing resolution in pixels per degree. We use the
Mannos-Sakrison contrast sensitivity function~\cite{mannos1974},
\begin{equation}
w_k^{\mathrm{CSF}}\propto
(0.0192+0.114vf_k)\exp[-(0.114vf_k)^{1.1}],
\end{equation}
normalize it so that $\sum_kw_k^{\mathrm{CSF}}=n$, and define
$\bar L^{\mathrm{CSF}}=n^{-1}\sum_kw_k^{\mathrm{CSF}}L_k$.

At the operating points of Table~\ref{tab:main_results}, the rankings for
$v\in\{16,32,64\}$\,px/deg agree closely with the unweighted ranking
($\rho_s=0.991$ for all three settings). Applying the same weights inside the
natural-image alignment statistic leaves the eight-NIC correlations unchanged
($\rho_s=-0.98$ with PSNR and $0.79$ with LPIPS). The tested weighting thus
rescales the profiles without reordering the codecs.

\subsection{Instability Sweeps and Gallery}
\label{sec:supp_x1_gallery}

\begin{figure*}[t]
  \centering
  \includegraphics[width=0.72\textwidth]{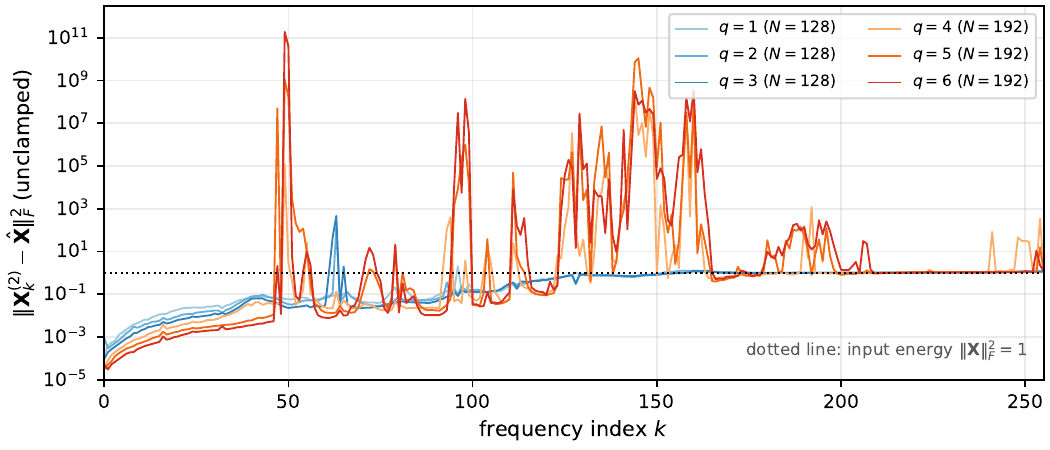}
  \caption{Unclamped squared reconstruction error on the
  $\mathbf{X}_k^{(2)}$ sweep ($n{=}256$, peak $0.45$) for the six released
  Cheng2020-Anchor checkpoints. The $N{=}128$, $q{=}1$-$3$ checkpoints have
  maximum errors below $500$; the $N{=}192$, $q{=}4$-$6$ checkpoints reach
  $1.9{\times}10^{11}$ at isolated indices (dotted line: unit input
  energy).}
  \label{sf:fig:instability}
\end{figure*}

\begin{figure*}[t]
  \centering
  \includegraphics[width=\textwidth]{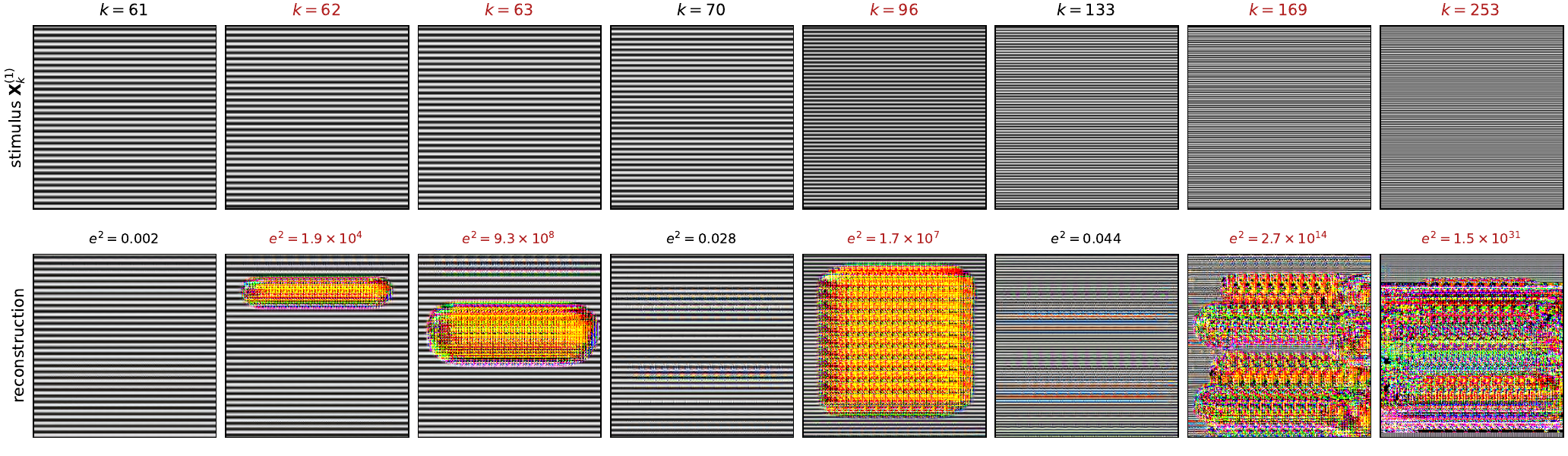}
  \caption{Instability of the grating probe $\mathbf{X}^{(1)}_k$
  ($n{=}256$, peak amplitude $0.45$): selected stimuli and their
  Cheng2020-Anchor $q{=}6$ reconstructions (clamped to $[0,1]$ for display).
  Red titles mark frequencies whose unclamped squared error exceeds $10$;
  artifacts range from a localized chromatic region ($k{=}62$) to global
  collapse ($k{=}253$), with interleaved stable indices.}
  \label{sfig:x1gallery}
\end{figure*}

Figure~\ref{sf:fig:instability} shows the unclamped $\mathbf{X}_k^{(2)}$
sweep underlying Sec.~\ref{sf:sub:instability}.
Complementing it, we sweep the grating stimuli $\mathbf{X}_k^{(1)}$ through
Cheng2020-Anchor without output clamping ($n{=}256$, centered embedding
$x\mapsto\frac12+sx$ with $s=an/\sqrt{2}$, i.e.\ peak amplitude
$a\in\{0.225,0.45\}$, matched to the half/full-contrast
$\mathbf{X}_k^{(2)}$ sweeps).
The gratings produce extreme pre-clamp errors at more frequencies than the
two-dimensional stimuli: $141$/$145$/$147$ of the $256$ frequencies exceed
$e^2{>}10$ at $q{=}4/5/6$ at full
amplitude ($36$/$24$/$49$ at half), and the unclamped error spans roughly
$35$ orders of magnitude across the sweep;
Fig.~\ref{sfig:x1gallery} shows stimulus/reconstruction pairs.

Divergence is not a smooth function of frequency: $k{=}61$ is
reconstructed near-perfectly while the neighboring $k{=}62$ and $k{=}63$
diverge, and stable indices ($k{=}70$, $k{=}133$) survive inside otherwise
unstable bands.
At moderate divergence the artifact is a localized chromatic region; at
extreme divergence ($k{=}169$, $k{=}253$) the reconstruction is corrupted
across the full image.
Because the divergent set shrinks at half amplitude, reported instability
sets must include the probe amplitude.

\subsection{Alignment Behavior Across Added Distortions}
\label{sec:supp_distcons}

We apply four additional distortions: Gaussian noise ($\sigma{\in}\{0.01,\dots,0.2\}$), pure noise, bit-depth quantization ($2$-$6$~bits), and JPEG re-compression ($q{\in}\{10,\dots,70\}$) to check whether the $L_k$-weighted statistic remains monotonic. Since $L_k$ is fixed per codec, only $\rho(f)$ changes. The evaluated distortions follow a broadly monotonic $\tilde{\mathcal{L}}$-PSNR trend; an unweighted $\bar{\rho}$ correlates marginally better ($\rho_s\,{=}{-}0.964$ vs.~$-0.945$, pooled over the eight NIC codecs), so the leakage weighting does not improve the prediction here. We report $\tilde{\mathcal{L}}/\bar{\rho}$ as a descriptive spectral-shape summary, not a predictive metric.

\section{Fine-Tuning Diagnostics and Ablations}

\begin{figure}[t]
  \centering
  \includegraphics[width=\columnwidth]{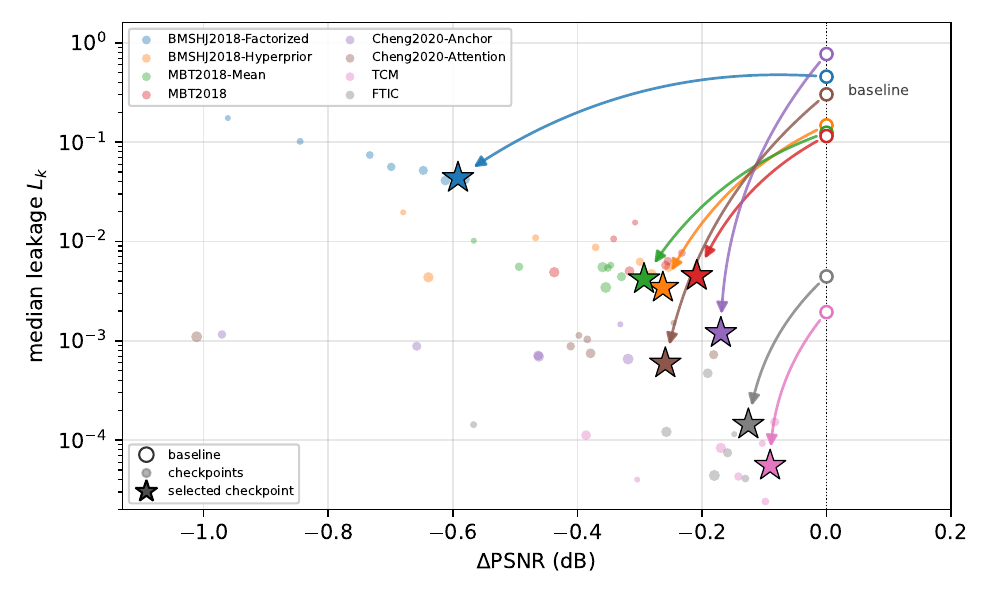}
  \caption{Leakage-regularized decoder fine-tuning trajectories at the
  $\lambda_{\mathrm{freq}}$ used in Table~\ref{new:tab:ft_joint}: median
  DCT-basis leakage ($256{\times}256$) vs.\ $\Delta$PSNR on the Kodak images.
  Small circles show checkpoints evaluated every 250 steps; arrows connect
  each pretrained~decoder to its selected checkpoint (star). For the seven
  codecs accepted under the $-0.1$\,dB validation limit, leakage decreases
  by $94.7$-$99.8\%$ and Kodak PSNR decreases by at most $0.29$\,dB.
  BMSHJ2018-Factorized reduces leakage by $90\%$, but has $\Delta\mathrm{PSNR}_{\mathrm{val}}{=}-0.416$\,dB.}
  \label{new:fig:pareto}
\end{figure}

Figure~\ref{new:fig:pareto} shows the leakage-PSNR selection trajectories at
the $\lambda_{\mathrm{freq}}$ of Table~\ref{new:tab:ft_joint}.
Table~\ref{new:tab:ft_basis} reports the basis-only negative control.

\begin{table}[t]
  \centering
  \caption{Basis-only decoder fine-tuning as a negative control. The decoder
  is optimized on one fixed DCT-basis input ($128{\times}128$; TCM and FTIC:
  $256{\times}256$) without natural images. $L^{\mathrm{med}}_{\mathrm{train}}$ is
  the median leakage on that same training probe; $\Delta$PSNR is fine-tuned minus
  pretrained performance on held-out Kodak images. The bitrate is unchanged
  because the encoder and entropy model remain frozen.}
  \label{new:tab:ft_basis}
  \smallskip
  \setlength{\tabcolsep}{4pt}
  \scriptsize
  \begin{tabular}{lcrrr}
\toprule
Codec & DCT size & $L^{\mathrm{med}}_{\mathrm{train}}$ before & $L^{\mathrm{med}}_{\mathrm{train}}$ after & $\Delta$PSNR (dB) \\
\midrule
BMSHJ2018-Fact. & $128{\times}128$ & 0.445 & 0.008 & -22.0 \\
BMSHJ2018-Hyp. & $128{\times}128$ & 0.163 & 0.003 & -13.9 \\
MBT2018-Mean & $128{\times}128$ & 0.148 & 0.002 & -16.5 \\
MBT2018 & $128{\times}128$ & 0.125 & 0.002 & -12.0 \\
Cheng2020-Anchor & $128{\times}128$ & 0.287 & $<0.001$ & -10.6 \\
Cheng2020-Attn. & $128{\times}128$ & 0.269 & $<0.001$ & -11.5 \\
TCM & $256{\times}256$ & 0.002 & $<0.001$ & -1.7 \\
FTIC & $256{\times}256$ & 0.004 & $<0.001$ & -1.9 \\
\bottomrule
\end{tabular}

\end{table}

\section{Selected Natural-Image Examples}

\subsection{Failure and Contrast Cases at Nearby Bitrates}
\label{sec:supp_gallery}

\begin{figure*}[p]
  \centering
  \includegraphics[width=0.82\textwidth]{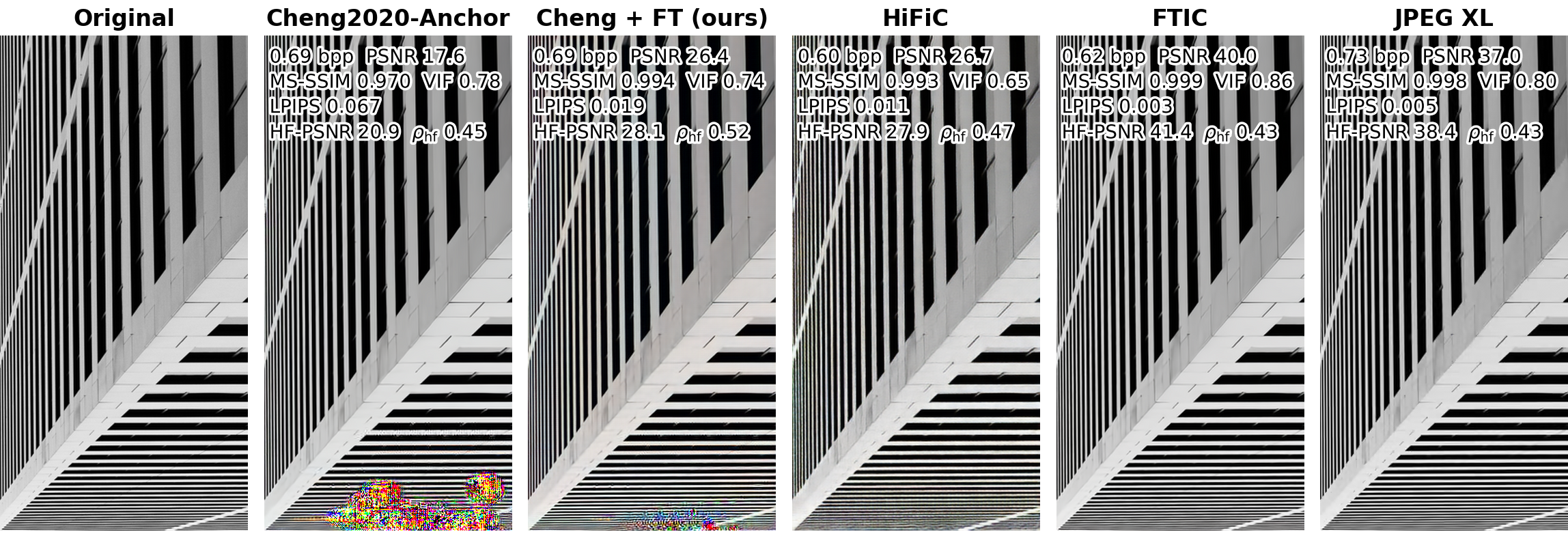}\\[2pt]
  \includegraphics[width=0.82\textwidth]{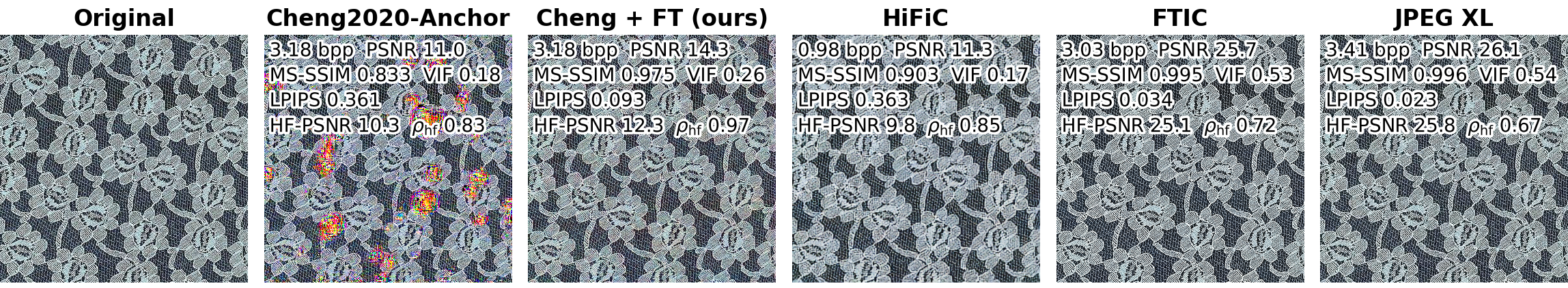}\\[2pt]
  \includegraphics[width=0.82\textwidth]{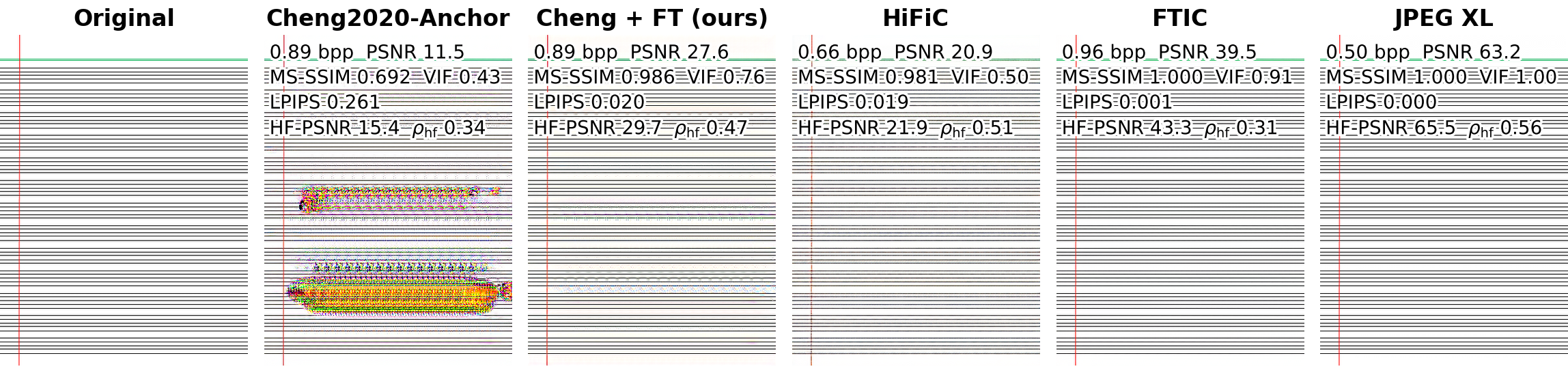}
  \caption{Affected near-periodic cases: Urban100 facade, DTD lace, and DTD
  parallel lines. Cheng2020-Anchor ($q{=}6$) produces localized chromatic
  artifacts; the fine-tuned decoder processes the same bitstream. FTIC and
  JPEG~XL are shown at the available settings nearest the Cheng bitrate.}
  \label{sfig:gallery1}
\end{figure*}

\begin{figure*}[p]
  \centering
  \includegraphics[width=0.82\textwidth]{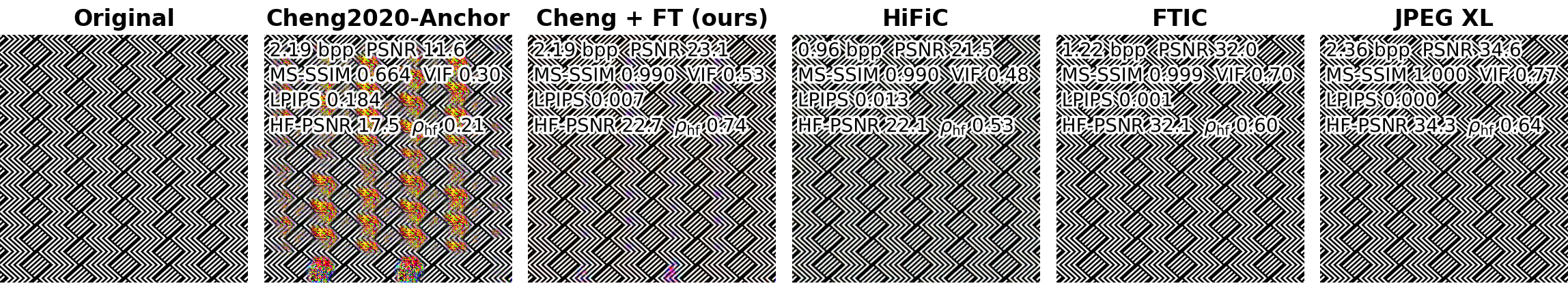}\\[2pt]
  \includegraphics[width=0.82\textwidth]{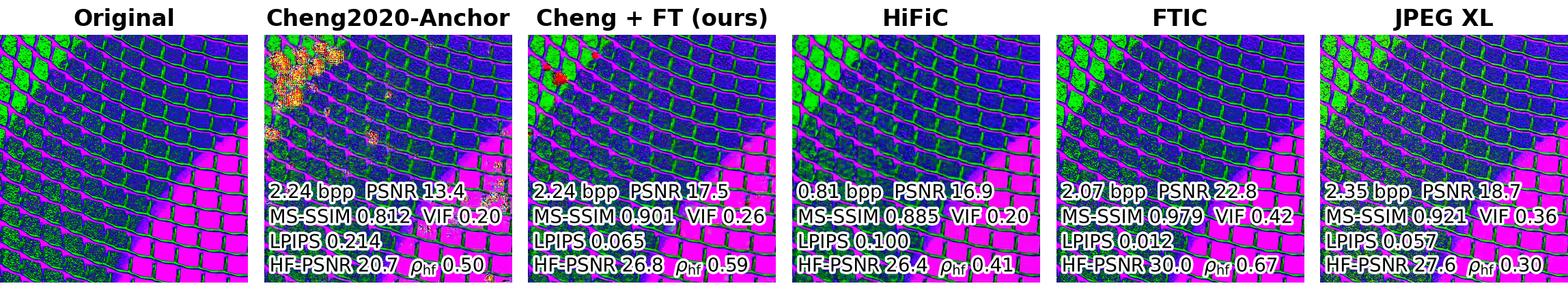}\\[2pt]
  \includegraphics[width=0.82\textwidth]{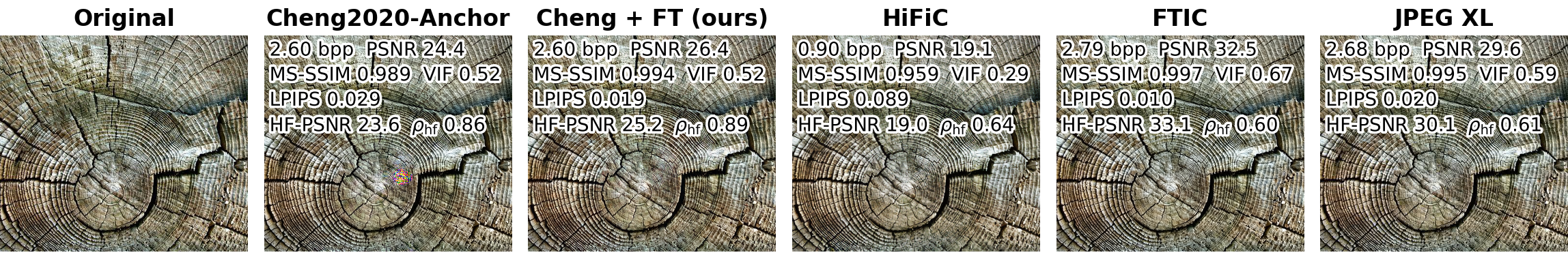}
  \caption{Two affected periodic textures (DTD zigzagged and meshed) and one
  minimally affected irregular texture (DTD cracked). The final row is a
  contrast case: high-frequency content alone does not imply a severe failure.
  Layout and overlays follow Fig.~\ref{sfig:gallery1}.}
  \label{sfig:gallery2}
\end{figure*}

Figures~\ref{sfig:gallery1}-\ref{sfig:gallery2} show six cases retained after
screening Urban100 and the Describable Textures Dataset (5{,}740 images in
total) with the benchmark pipeline. Five rows cover distinct pronounced
failures (facade slats, lace, parallel lines, zigzags, and a mesh); the
irregular cracked texture is included as a minimally affected contrast. The
selection illustrates the range of observed behavior, not its prevalence.
Each row shows the original crop,
Cheng2020-Anchor at $q{=}6$, the same model after the decoder fine-tuning of
Sec.~\ref{new:sub:ft_joint}, HiFiC at its highest released rate, and FTIC and
JPEG~XL at the settings nearest the Cheng bitrate. Crops are deterministic:
the $512{\times}512$ window of maximum high-frequency energy, or the largest
$64$-multiple center crop for smaller images. Overlaid on each reconstruction:
bitrate, PSNR, MS-SSIM, VIF, LPIPS, HF-PSNR (PSNR restricted to the high
band, radial frequency above half-Nyquist), and $\rho_{\mathrm{hf}}$ (the
distortion fraction $\rho(f)$ of Sec.~\ref{sec:weighted_leakage} integrated
over the high band).

On the affected near-periodic examples, Cheng2020-Anchor produces localized
chromatic artifacts and severe quality loss (PSNR $11$-$18$\,dB on the lace,
line, and facade rows). These artifacts are visually related to the
single-frequency failure of Sec.~\ref{sf:sub:instability}, but the experiment
does not establish a shared mechanism. The fine-tuned decoder removes most of
the visible artifacts from the same bitstream, recovering $3$-$16$\,dB on
the affected rows (facade
$17.6\!\to\!26.4$\,dB, lined $11.5\!\to\!27.6$\,dB) at zero rate cost. FTIC
and JPEG~XL show no comparable artifacts at the nearest available bitrates.
HiFiC is not rate-matched in every row; its comparison is therefore
descriptive.

\section{Reproducibility}
\label{sec:supp_repro}

\begin{sloppypar}
Source code, evaluation scripts, pre-computed results, notebooks that
regenerate every figure, and an installable package are available at
\url{https://github.com/nkalmykovsk/dct_benchmark_nic}, with pinned dependency
versions. This section states the settings needed to read the results; the
repository holds the remaining implementation detail, including the mapping
from each figure and table to the script that produces it.
\end{sloppypar}

\textbf{Environment.}
Benchmark experiments use Python~3.11 and PyTorch~2.9 on one NVIDIA A100
(40~GB); extended experiments use Python~3.9 and PyTorch~2.5.1 on one NVIDIA
H100 (80~GB). Neural codecs are taken from CompressAI~1.2.8 (BMSHJ and MBT
families, Cheng2020-Anchor, Cheng2020-Attention) and from the reference
implementations of TCM and FTIC; classical codecs from Pillow (JPEG, WebP) and
imagecodecs (JPEG~XL); metrics from the reference LPIPS and MS-SSIM
implementations.

\textbf{Determinism.}
The basis $D{=}\mathrm{DCT}(I_n)$ is deterministic, and all codec weights are
frozen pre-trained checkpoints evaluated in inference mode. The basis-only
control was repeated with three seeds (leakage standard deviation
${<}0.001$); joint fine-tuning uses a single fixed seed, with no multi-seed
stability claim.

\textbf{Codec configurations.}
CompressAI models and FTIC are evaluated at quality levels
$q{\in}\{1,\dots,6\}$, TCM at $\lambda{\in}\{0.0025, 0.05\}$ with
$p{\in}\{64,128\}$, JPEG and WebP at qualities $\{20,40,55,70,85,95\}$, and
JPEG~XL at distances $\{4.0,2.0,1.0,0.6,0.3,0.1\}$. DCT bases use
$n{\in}\{64,128,256,512,1024\}$, restricted to $n{\geq}256$ for FTIC and TCM.

\textbf{Bitrate.}
For the neural codecs, bpp is the entropy-model estimate, i.e.\ the
cross-entropy of the quantized latents under the learned likelihoods, which is
the standard CompressAI convention and a tight proxy for the arithmetic-coded
length. For JPEG, WebP, and JPEG~XL, bpp is the encoded file size divided by
the pixel count. Rate-matched comparisons therefore compare an estimated rate
against a measured one.

\textbf{Table~\ref{tab:main_results}.}
115 images: 24~Kodak at native resolution, 41~CLIC~Pro validation, and
50~DIV2K (evenly subsampled), the latter two center-cropped to
${\leq}768$\,px. NICs at fixed quality $q{=}6$; classical codecs at the
quality whose mean bitrate is nearest $0.9$\,bpp per dataset.
$L_k$ is computed at the same settings on a $512{\times}512$ basis; $N_b{=}512$ radial bins.
PSNR, MS-SSIM, LPIPS computed on RGB; Spearman correlations over 8~NICs.

\textbf{Numerical caveat.}
TF32 must be disabled on Ampere and Hopper GPUs: with TF32 enabled, the
MS-SSIM implementation returns invalid values ($>1$).

\subsection{Fine-Tuning Protocols}
\label{supp:ft_protocols}

\textbf{Basis-only controls.}
The basis-only experiment of main-paper Fig.~\ref{fig:freq_analysis}(c)
updates only the synthesis transform $g_s$ with Adam
($\mathrm{lr}{=}10^{-4}$) for 900 steps on one DCT-basis image
($128{\times}128$). The objective sets
$\mathcal L_{\mathrm{dist}}{=}0$ and $\lambda_{\mathrm{freq}}{=}1$ and uses
$p{=}2$; thus, the normalized weights
$w_k{\propto}(k/(n-1))^2{+}0.2$ are zero for
$k{<}\lfloor0.33n\rfloor$. It also includes the purity-deficit and low-band
MSE terms of Sec.~\ref{sec:metric_loss}. Table~\ref{new:tab:ft_basis}
uses the same decoder-only objective for $1000$ steps and three seeds on a
$128{\times}128$ basis (TCM and FTIC: $256{\times}256$).

\textbf{Table~\ref{new:tab:ft_joint}: data and updates.}
Eight DIV2K images are the held-out validation set; the fine-tuning set is the
first 400 remaining images in sorted order. Each step samples four images with
replacement and draws one random $256{\times}256$ crop from each (8000 crops
per 2000-step run) and processes one fixed $256{\times}256$ DCT basis
($\mathcal L_{\mathrm{freq}}$ with $p{=}3$, both guards). All runs use Adam
($10^{-4}$), gradient-norm clipping at 1, and a single fixed seed;
leakage-regularized and MSE-only runs share the fine-tuning set but use
independent minibatch sequences. The MSE-only control sets
$\lambda_{\mathrm{freq}}{=}0$ with both guards disabled.

\textbf{Checkpoint grid.}
Checkpoints are evaluated at steps $\{0,250,\ldots,2000\}$ and selected by
the validation rule of Sec.~\ref{new:sub:ft_joint}. We evaluate
$\lambda_{\mathrm{freq}}{=}0.03$ for all models and additionally $0.01$ and
$0.003$ for the four BMSHJ/MBT models; BMSHJ2018-Factorized has no
nonzero-step checkpoint that passes the gate, and
Table~\ref{new:tab:ft_joint} reports its $\lambda_{\mathrm{freq}}{=}0.003$,
step-2000 endpoint for completeness. CLIC is used for the held-out example
in Fig.~\ref{new:fig:ft_natural} and Figs.~\ref{sfig:gallery1}-\ref{sfig:gallery2}.

\subsection{Probe Amplitudes}
\label{supp:amplitudes}

\looseness=-1
The two amplitude conventions are not interchangeable
(Sec.~\ref{sec:supp_singlefreq}). The single-frequency suite embeds
$\mathbf{X}^{(1)}_k$ and $\mathbf{X}^{(2)}_k$ by $x\mapsto\frac12+sx$ over all
$k$, with $s{=}0.225n$ for $\mathbf{X}^{(2)}$ and $s{=}0.45\sqrt{n}$ for
$\mathbf{X}^{(1)}$ at $n{=}256$, clamped to $[0,1]$ unless stated. The
grating-instability runs (Fig.~\ref{sf:fig:artifact_visual},
Fig.~\ref{sfig:x1gallery}) instead use $s{=}an/\sqrt{2}$ with peak amplitude
$a\in\{0.225,0.45\}$ and no clamping; reported instability sets are valid only
for the stated amplitude. The profile sweep covers 124 configurations.

\end{document}